\documentclass[12pt]{article}
\usepackage[margin=1in]{geometry}
\usepackage{setspace}
\usepackage{amsfonts}       
\usepackage{amsthm}
\usepackage{dsfont}     
\usepackage{booktabs}
\usepackage{float}
\usepackage{graphicx}
\usepackage{hyperref}
\usepackage{mathtools}
\usepackage{microtype}
\usepackage{subfig}
\usepackage{thm-restate}
\usepackage{algorithm, algorithmic}
\usepackage{color}

\newtheorem{lemma}{Lemma}
\newtheorem{theorem}{Theorem}

\newcommand{\field}[1]{\mathbb{#1}}
\def \E{\field{E}}

\def \P{\field{P}}
\def\R{\mathbb{R}}
\def\Z{\mathbb{Z}}
\def\eps{\epsilon}

\DeclareMathOperator\erf{erf}

\newcommand{\aggregatecounts}{\texttt{Aggregate-Counts}}
\newcommand{\hammingdistancesampler}{\texttt{Hamming-Distance-Sampler}}
\newcommand{\countsvector}{C^k_i}
\newcommand{\countsvectorquery}{C^k_q}
\newcommand{\countsvectornotable}{C_i}
\newcommand{\countsvectornotablej}{C_j}
\newcommand{\countsvectornotablequery}{C_q}
\newcommand{\countsmatrix}{M^k_i}
\newcommand{\countsmatrixnotable}{M_i}
\newcommand{\neighbors}{\mathcal{N}_i}
\newcommand{\queryhash}{{\bf{0}}_t}

\newcommand{\buckets}{\mathcal{B}^k_q}

\newcommand{\bucketsMulti}{\mathcal{B}^k_m(q)}
\newcommand{\projectionsMulti}{\mathcal{R}^k_q}

\newcommand{\firstEstimator}{\emph{LSH Count}}
\newcommand{\secondEstimator}{\emph{Multi-Probe Count}}

\title{Local Density Estimation in High Dimensions}

\author{
	Xian Wu \\
	Stanford University \\ 
	\texttt{xwu20@stanford.edu}
	\and
	Moses Charikar \\
	Stanford University \\
	\texttt{moses@cs.stanford.edu}
	\and
	Vishnu Natchu \\
	Laserlike Inc \\
	\texttt{vishnu@laserlike.com}
	}

\date{}
\begin{document}

\maketitle

\begin{abstract}
An important question that arises in the study of high dimensional vector representations learned from data is: given a set $\mathcal{D}$ of vectors and a query $q$, estimate the number of points within a specified distance threshold of $q$. We develop two estimators, \firstEstimator\space and \secondEstimator\space that use locality sensitive hashing to preprocess the data to accurately and efficiently estimate the answers to such questions via importance sampling. A key innovation is the ability to maintain a small number of hash tables via preprocessing data structures and algorithms that sample from multiple buckets in each hash table. We give bounds on the space requirements and sample complexity of our schemes, and demonstrate their effectiveness in experiments on a standard word embedding dataset.
\end{abstract}

\section{Introduction}
\label{intro}

In this work, we study a basic question that arises in the study of high dimensional vector representations: given a dataset $\mathcal{D}$ of vectors and a query $q$, estimate the number of points within a specified distance threshold of $q$.
Such density estimates are important building blocks in non-parametric clustering, determining the popularity of topics, search and recommendation systems, the analysis of the neighborhoods of nodes in social networks, and in outlier detection, where geometric representations of data are frequently used. Yet for high dimensional datasets, we still lack simple, practical, experimentally verified and theoretically justified solutions to tackle this question.

Our questions have been studied in the context of \emph{spherical range counting}. One class of solution methods arising in the computational geometry literature, such as hierarchical splitting via trees, \cite{spherical-range} have performance guarantees that depend exponentially on dimension. These are unsuitable for the higher dimensional models that machine learning methods are increasingly shifting towards e.g. word embeddings \cite{pennington2014glove,Mikolov:2013aa} and graph embeddings \cite{perozzi2014deepwalk,tang2015line,cao2015grarep,grover2016node2vec,yang2016revisiting,wang2017community,hamilton2017inductive}. Over-parameterized models are oftentimes easier to train \cite{NIPS2014_5267}, and perform just as well, if not better \cite{Zhang:2016aa}. Word embeddings is one example where rigorous evaluation has shown increased performance with higher dimensionality \cite{Melamud:2016aa} \cite{siwei}.

In this paper, we develop two estimation schemes, \firstEstimator\space and \secondEstimator, for high dimensional datasets to count the number of elements around a query that are in a given radius of cosine similarity. Angular distance, which corresponds to Euclidean distance for data points on the unit sphere is commonly used in applications related to word and document embeddings, and image and video search \cite{produce-quantization} \cite{Huang:2012:IWR:2390524.2390645}. Brute force search requires a linear scan over the entire dataset, which is prohibitively expensive. Our approach uses indexing and search via locality sensitive hashing (LSH) functions in order to estimate the size of the neighborhood in a more efficient manner than retrieving the neighbors within the given radius of similarity. 

Our approach improves upon the storage and sample complexities of previous methods using a combination of extracting information from multiple buckets per table (hence reducing table complexity) and importance sampling (hence reducing sample complexity). As we show in our experimental study on GLOVE embeddings, our estimates of the number of elements that are 60 degrees from a query $q$ (which corresponds to synonyms and/or related words to $q$ in the English vocabulary), achieve multiple orders of magnitude improved accuracy over competing methods, subject to reasonable and practical resource constraints. Our theoretical analysis develops a rigorous understanding of our technique and offers practitioners further insight on optimizing our solution method for their particular datasets.

\section{Related Literature}
\label{related_literature}
The biggest challenge in high dimensional statistics is the curse of dimensionality. There are two major perspectives on high dimensional data; the first philosophy is that most datasets are inherently and latently low rank, so techniques that depend on dimension can be fruitfully used, as long as one first finds the low rank approximation to the high dimensional data. The second approach is to develop dimension-free methods that can directly and effectively work with high dimensional data. In the dimension-free toolbox, Locality Sensitive Hashing (LSH) for Approximate Nearest Neighbor Search (ANN) was introduced in \cite{Indyk:1998:ANN:276698.276876}. This influential work started a long line of research to develop and analyze better Locality Sensitive Hashing schemes, see \cite{terasawa2007spherical}\cite{NIPS2015_5893}\cite{kennedy2016fast}\cite{Andoni:2015:ODH:2746539.2746553}\cite{datar2004locality} and study their limits, see \cite{andoni2015tight}\cite{o2014optimal}\cite{motwani2006lower}. 

More recently, researchers have begun leveraging LSH techniques to solve problems beyond ANN, extending their domain to applications around density estimation for high-dimensional models. For example \cite{Aumuller:2017aa} generalizes nearest neighbor LSH hash functions to be sensitive to custom distance ranges. \cite{ahle2017parameter} builds many different parameterized versions of the prototypical LSH hash tables and adaptively probes them for spherical range reporting. 

The closest works to ours in terms of proposed solution method that we are aware of is that of \cite{Spring:2017aa}, which gives an LSH based estimator to compute the partition function of a log-linear model. \cite{mc-ps} adapts LSH to solve a class of kernel density estimation problems. Our problem can be viewed as a kernel density estimation problem for a specific kernel function that has value $1$ for pairs of points within the required angle range of interest and $0$ outside.
However the analysis of \cite{mc-ps} does not apply to our setting because they need a {scale free} hash function (with collision probabilities related to the kernel value) and there is no such function for our 0-1 kernel.

These two works are similar to ours in their application of importance sampling; the major difference is that the techniques in both works leverage only one hash bucket per table, and hence requires a large number of tables for an accurate estimate. The biggest drawback to these works is the very high storage (hash tables) and query complexities -- their techniques, as presented, are impractical for adoption. 

One practical, storage-efficient scheme originally conceived for LSH for ANN is multi-probe \cite{lv2007multi}, the technique of probing from multiple buckets in a hash table based on the success probability of the bucket containing a nearest neighbor. This technique builds on the entropy-based probing scheme introduced by \cite{panigrahy2006entropy}. Our technique also leverages multiple buckets per table, and the biggest difference is that we importance-weigh our samples. In our experimental results section, we compare the efficacy of our algorithms against multi-probe and the estimator developed in \cite{Spring:2017aa}.

\section{Problem Formulation and Approach Overview}
Given a dataset $\mathcal{D}$ of vectors $v_1, \ldots v_n \in \R^d$ on the unit sphere, a query $q \in \R^d$ also on the unit sphere, and a range of angles of interest $\mathcal{A}$, for example 0-60 degrees, how many elements $v$ in $\mathcal{D}$ are such that the angle between $q$ and $v$, denoted $\theta_{qv}$, are within range $\mathcal{A}$? We use $\mathcal{A}_q$ to denote the set of data vectors $v$ that are within angle $\mathcal{A}$ to $q$ (that have angular distance to query $q$ that is in the range of interest $\mathcal{A}$). Our goal is to preprocess $\mathcal{D}$ in order to estimate the cardinality of this set, denoted $|\mathcal{A}_q|$, efficiently for any given $q$. 

One final note is that our schemes are conceptualized using bit-wise LSH functions; functions that hash vectors to 0-1 bits, and where the hamming distance between the hash sequences of two data points captures information about their angular distance. For their simplicity, easy implementation, and high performance in practice, bit hashes such as hyperplane LSH \cite{Charikar:2002:SET:509907.509965} are the standard hash functions used in practice for angular distance \cite{NIPS2015_5893}. Our technique and results can be extended for other hash functions; however, we will use hamming distance and other implementation details specific to bit-wise LSH functions in this work. 

\subsection{Approach Overview}
We introduce our two schemes, \firstEstimator\space and \secondEstimator. While the finer details of the algorithms are different, they share in two main ideas: importance weighting the samples and probing from multiple buckets from the hash table. They consist of two steps, a preprocessing step that applies locality sensitive hash functions to our dataset to produce hash tables. After this preprocessing step, we sample from our hash tables to produce our final estimate. We first offer an intuitive introduction to the high-level workings of our algorithms.

Importance sampling concentrates the elements of interest in our overall dataset into a few buckets that we can easily sample from to produce our estimate. In order to compensate for the concentrated sampling, we adjust the value of each sample by the inverse of the probability that the sample lands in the target buckets. 

Our techniques rely on the key insight that LSH functions can effectively implement both of these objectives. Using LSH functions to index our dataset ensures that for a given query $q$, elements that are close to $q$ in angular distance have a comparative higher probability of hashing to $q$'s bucket and to buckets that are of small hamming distance to $q$'s bucket, thereby concentrating the elements of interest into certain buckets that we can selectively sample from. 

Additionally, the hamming distance collision probabilities for bit-wise LSH functions are well expressed in terms of angular distance. Consider random hyperplane LSH \cite{Charikar:2002:SET:509907.509965}, where each hash vector is chosen from the $d$-dimensional Gaussian distribution (each coordinate is drawn from the 1-dimensional Gaussian distribution). Each hash vector $r$ contributes one bit to the hash sequence of a data point $v$, based on the rule: 

\[
h_r(v) = \begin{cases} 
      0 & \text{ if } r \cdot v \leq 0 \\
     1 & \text{ otherwise. }
   \end{cases}
\]
It is well-known that for any particular hamming distance $i$, and any data point $x$, 
\begin{equation}\label{collision_prob_bit}
\P(d_{qx} = i | \theta_{qx}) = {t \choose i} \left(1 - \frac{\theta_{qx}}{\pi}\right)^{t-i} \left(\frac{\theta_{qx}}{\pi}\right)^i,
\end{equation}
where $d_{qx}$ is the hamming distance between the hash for query $q$ and the hash for data vector $x$, $\theta_{qx}$ denotes the angle between the 2 vectors, and $t$ is the total number of bits in the hash sequence. 

In Appendix \ref{appendix_C} we extend this result to give a finer-grained analysis of the probability of $x$ hashing to a specific bucket address (not just the set of buckets of hamming distance $i$), given the exact values of the query's projection onto each of the random hyperplanes. 

Thus, the choice of $t$ affects the sensitivity of the LSH scheme -- the correlation between the hamming distances of two hash sequences and the angle between the two underlying data points. Moreover, depending on the design choice for $t$, the set of buckets or the set hamming distances $\mathcal{I}$ that contains most of the probability mass for collision with elements of angular distance in range $\mathcal{A}$ is different. This is also a consideration in our sampling scheme; we want to sample from buckets that have a high probability of containing elements that are within angle $\mathcal{A}$ of $q$. 

In this paper, we introduce two estimation schemes, \firstEstimator\space and \secondEstimator. \firstEstimator\space samples from elements at hamming distances $\mathcal{I}$ from the query and uses Equation \ref{collision_prob_bit} to weigh the samples. \secondEstimator\space samples from specific buckets and importance weights according to fine-grained bucket probability of containing an interesting element. \secondEstimator\space uses collision probabilities that we develop in Appendix \ref{appendix_C}. We now go into more detail about each of the two schemes. 

\subsection{\firstEstimator}
\firstEstimator\space picks elements over $K$ hash tables from buckets that are at hamming distance $\mathcal{I}$ to the query, where $\mathcal{I}$ is tuned to $\mathcal{A}$. Given a sample, $x$, we compute the angular distance $\theta_{qx} = \cos^{-1} (q \cdot x)$. Let $p(x) = \P(d_{qx} \in \mathcal{I} | \theta_{qx})$, the collision probability that $x$ lands in a bucket that is hamming distance $\mathcal{I}$ from $q$ over the random choice of hash functions. 

We define a random variable $Z$ as a function of sample $x$ as follows:
\begin{equation}\label{eq:score_function}
Z = \begin{cases} 
      \frac{\sum^K_{k=1}\countsvectorquery({\mathcal{I}})}{K \cdot p(x)} & \text{ if } \theta_{qx} \in \mathcal{A} \\
     0 & \text{ otherwise. }
   \end{cases}
\end{equation}
where $\countsvectorquery({\mathcal{I}})$ is the total number of elements in buckets of hamming distance $\mathcal{I}$ from $q$'s bucket in table $k$. 

We take $S$ samples and construct $Z_1, Z_2, \ldots Z_S$. We report $\frac{\sum_{i=1}^S Z_i}{S}$ as our estimate for $|\mathcal{A}_q|$. 

We establish the following theoretical bounds on the storage and sample complexity of \firstEstimator\space in order to achieve a $(1\pm\eps)$-approximation to the true count with high probability. 

\begin{restatable}[\firstEstimator]{theorem}{mainresult}
\label{thm:mainresult}
For a given angular distance range of interest $\mathcal{A}$ and a given query $q$, with probability $1-\delta$, our estimator returns a $(1\pm \eps)$-approximation to $|\mathcal{A}_q|$, the true number of elements within angle $\mathcal{A}$ to $q$ using $O\left(\frac{1}{\eps^2 \min\limits_{x \in \mathcal{A}_q} p(x)} \log(\frac{1}{\delta})\right)$ tables and $O\left(\frac{\E(\countsvectornotablequery({\mathcal{I}}))}{\eps^2|\mathcal{A}_q| \cdot \min\limits_{x \in \mathcal{A}_q}p(x)}\log(\frac{1}{\delta})\right)$ samples. 
\end{restatable}

To help the reader digest this result, we briefly compare this statement to the sample complexity of naive random sampling. It can be shown through a standard Bernoulli-Chernoff argument that the sample complexity for random sampling is $O\left(\frac{n}{|\mathcal{A}_q| \eps^2} \ln \left(\frac{1}{\delta}\right)\right)$, where $\frac{n}{|\mathcal{A}_q|}$ is the inverse proportion of elements of interest in the overall population. Intuitively this says that you need to take more random samples if $|\mathcal{A}_q|$ is very small compared to $n$. 

Our sample complexity replaces the $\frac{n}{|\mathcal{A}_q|}$ term with $\frac{\E(\countsvectornotablequery({\mathcal{I}}))}{|\mathcal{A}_q| \cdot \min\limits_{x \in \mathcal{A}_q}p(x)}$, where $|\mathcal{A}_q| \cdot \min\limits_{x \in \mathcal{A}_q} p(x)$ is a measure of the expected number of elements from the set of interest $\mathcal{A}_q$ that will land in hamming distance $\mathcal{I}$ to $q$, and $\E(\countsvectornotablequery({\mathcal{I}}))$ is the expected size of the overall sampling pool of elements in hamming distance $\mathcal{I}$. This ratio of expectations seems intuitive -- one would expect to get such an expression if our scheme took one sample per table. Surprisingly, we achieve this same type of sample complexity bound  while sampling from relatively few hash tables.

Just like random sampling, our sample complexity bound is also based on the proportion of elements of interest in hamming distance $\mathcal{I}$ to the total number of elements in hamming distance $\mathcal{I}$. However, it is easy to see that applying LSH to our dataset will increase this proportion to yield a smaller sample complexity. We choose $\mathcal{I}$ so that $\min\limits_{x \in \mathcal{A}_q} p(x)$ is high (this probability can be high even for a small set of hamming distances $\mathcal{I}$, since $p(x)$ is the cumulative probability mass of $\mathcal{I}$ successes in $t$ trials, and binomial distributions in $t$ concentrate in an $O(\sqrt{t})$ sized interval around the mean), and $\E(\countsvectornotablequery({\mathcal{I}}))$ to be small (to filter out elements that are not interesting). 

There are certain tradeoffs to choosing $\mathcal{I}$. If more hamming distances are included in $\mathcal{I}$, then $\min\limits_{x \in \mathcal{A}_q}p(x)$ is higher, however, $\E(\countsvectornotablequery({\mathcal{I}}))$ is also larger. The optimal choice for $\mathcal{I}$ is to choose the hamming distances that substantially increase $\min\limits_{x \in \mathcal{A}_q}p(x)$ yet do not substantially increase $\E(\countsvectornotablequery({\mathcal{I}}))$ (so not too many uninteresting elements are infiltrating those buckets). 

\subsection{\secondEstimator}
The idea of \secondEstimator\space is to use the same importance weighing technique used in the \firstEstimator, however, instead of sampling from hamming distances $\mathcal{I}$, we search promising buckets according to the query-directed probing idea introduced in \cite{lv2007multi}. 

The query-directed probing idea is the following: instead of searching in buckets that are $0, 1, 2, \ldots$ hamming distances away from the query bucket, it is useful to take into account the value of the query projection onto the set of random hyperplanes that underlies the hash table. If a query had a strong projection onto a particular random hyperplane, it is more likely that a near neighbor to the query would share the same hash value as the query for that particular random hyperplane. When probing multiple buckets in a hash table, therefore, it could be better to first flip bits corresponding to hyperplanes that are near-orthogonal to the query. It may even be more fruitful to flip 2 ``weak" bits before flipping 1 ``strong" bit.

The query-directed probing idea relies heavily on computing fine-grained collision probabilities for each bucket in the table, given the value of the query's projection onto each random hyperplane underlying the hash table. We provide a calculation for random hyperplane LSH in Appendix \ref{appendix_C} that we use to implement \secondEstimator.

Let $\bucketsMulti$ denote the set of hash buckets to multi-probe from in the $k$-th hash table. Let $\projectionsMulti$ denote values of the projection of the query onto the random hyperplanes for table $k$. We then score each element found as: 
\begin{equation}\label{eq:score_function_multiprobe}
Z = \begin{cases} 
      \frac{1}{\sum\limits_{k \in K} p^k_m(x)} & \text{ if } \theta_{qx} \in \mathcal{A} \\
     0 & \text{ otherwise. }
   \end{cases}
\end{equation}
where $p^k_m(x)$ is the probability that $x$ is hashed to the set of buckets $\bucketsMulti$, that is $p^k_m(x) = \P(x \in \bucketsMulti | \theta_{qx}, \projectionsMulti)$. Appendix \ref{appendix_C} gives the formula for computing this probability for random hyperplane LSH. 

\secondEstimator\space inspects the elements in the chosen promising buckets, it does not sample, this is one key difference between \secondEstimator\space and \firstEstimator. We take all $S$ elements that we inspect, weighted as $Z_1, Z_2, \ldots Z_S$ and report $\sum^S_{i=1} Z_i$ as the final estimate. 

One last remark that we wish to make is that in Appendix \ref{appendix_C}, our calculation shows that $p^k_m(x)$ actually depends on $\theta_{qx}$, which is problematic for the reason that we use $p^k_m(\cdot)$ to rank the buckets to establish a probing order. When we are interested in a range of angles, say 0 to 60 degrees, one can use a representative angle, such as 45 degrees, to rank the buckets, or a few representative angles.

We establish the following theoretical bounds on the storage and sample complexity of \secondEstimator\space in order to achieve a $(1\pm\eps)$-approximation to the true count with high probability. 
\begin{restatable}[\secondEstimator]{theorem}{mainresultmulti}
\label{thm:mainresultmulti}
For a given angular distance range of interest $\mathcal{A}$ and a given query $q$, with probability $1-\delta$, \secondEstimator\space returns a $(1\pm \eps)$-approximation to $|\mathcal{A}_q|$, the true number of elements within angle $\mathcal{A}$ to $q$ when $\min_{x \in \mathcal{A}_q} \sum\limits_{k \in K} p^k_m(x) =O( \frac{1}{\eps^2}\log(\frac{1}{\delta}))$.
\end{restatable}

\subsection{Paper Overview}
In the rest of this paper, we explain our two estimators \firstEstimator\space and \secondEstimator\space further and present our experimental results. Our paper is organized into the following sections: 
\begin{itemize}
\item Sections \ref{sec:preprocessing} and \ref{sec:sampling} are devoted to developing \firstEstimator.
Section  \ref{sec:preprocessing} details some preprocessing implementations to enable \firstEstimator\space and gives provable bounds on the number of hash tables required for \firstEstimator. More implementation details for some preprocessing routines are deferred to Appendix \ref{appendix_A} and Appendix \ref{appendix_B}.
Section \ref{sec:sampling} presents the analysis for the sample complexity of \firstEstimator. 
\item Section \ref{sec:multi-probe_importance} presents the analysis for \secondEstimator. The collision probability calculations are derived in Appendix \ref{appendix_C}. This section gives provable bounds on the number of required hash tables. The analysis in this section is similar to the analysis previously developed for \firstEstimator\space in sections \ref{sec:preprocessing} and \ref{sec:sampling}.
\item Section \ref{sec:experiments} gives the computation results of our experiments on a standard word embedding dataset, compared to benchmark techniques of \cite{Spring:2017aa} and \cite{lv2007multi}. 

Implementation details for multi-probe \cite{lv2007multi}, specialized to our setting of random hyperplane projection and angular distance, are developed in Appendix \ref{appendix_C}.
\end{itemize}

\section{Preprocessing for \firstEstimator} \label{sec:preprocessing}

The preprocessing step contributes 3 key ingredients to \firstEstimator: 

\textbf{Hash Tables: } Given a family of bit-wise hash functions $\mathcal{H}$, define a function family $\mathcal{G} = \{ g : \mathcal{D} \rightarrow \{0,1\}^t\}$ such that $g(v) = (h_1(v), \ldots h_t(v))$, where $h_j \in \mathcal{H}$. To construct $K$ tables, we choose $K$ functions $g_1, g_2, \ldots g_K$ from $\mathcal{G}$ independently and uniformly at random. We store each $v \in \mathcal{D}$ in bucket $g_k(v)$ for $k = 1, 2 \ldots K$. This step sets up the hash tables that we will sample from in our scheme.

\textbf{Counts Vector: } We create a counts vector, denoted $\countsvector \in \R^{t+1}$ for each hash address $i^k$ for each table $k \in \{1, \ldots, K\}$, where $\countsvector (d)$ is the count of the total number of items in buckets that are at hamming distance $d=0, 1, \ldots t$ away from $i^k$ in table $k$.

\textbf{Sampler: } We create a sampler that given a separate hash address $i^k$ for each table $k \in \{1, \ldots, K\}$ and set of hamming distances $\mathcal{I}$, returns a data point uniformly at random from the union of elements that were hashed to buckets of hamming distance $\mathcal{I}$ from $i^k$ across the $K$ tables.

In the rest of this section, we describe in greater detail the 3 main components of the preprocessing step.

\subsection{Hash Tables}
\label{sec:hash_tables}

Setting up quality hash tables to enable accurate and efficient importance sampling is vital to \firstEstimator. Since we are importance sampling from buckets of hamming distance $\mathcal{I}$ across $K$ tables, we need to make enough tables to guarantee unbiasedness or near-unbiasedness for our sampling-based estimator; due to the variance of the randomly generated hash functions, if we make too few tables we may not find enough elements of interest contained in those tables within hamming distance $\mathcal{I}$. We want to characterize the bias of our importance sampling scheme in relation to the contents of the buckets of our hash tables. 

We let $\buckets(\mathcal{I})$ denote the set of hash buckets that are at hamming distance $\mathcal{I}$ from the hash address of query $q$ for table $k$. Next, we introduce an intermediate random variable: 
\[
W = \frac{1}{K} \sum\limits_{k=1}^K \sum_{x \in \mathcal{A}_q} \frac{\mathds{1}(x \in \buckets(\mathcal{I}))}{p(x)} ~,
\]
where $p(x) = \P(d_{qx} \in \mathcal{I} | \theta_{qx})$. 

$W$ is a random variable that represents the sum of the elements of interest $|\mathcal{A}_q|$ that are hashed to the buckets of sampling focus $\buckets(\mathcal{I})$, weighted by their probabilities $p(x)$. It is clear that once the set of hash functions is fixed, $W$ becomes deterministic. 

We first show that the random variable $Z$, as defined in Equation~\eqref{eq:score_function}, is an unbiased estimator. 

\begin{restatable}[Expectation of $Z$]{lemma}{expectationz}
\label{lem:expectation_z}
The expectation of $Z$ over the random choice of hash functions is $|\mathcal{A}_q|$, i.e. $\E(Z) = |\mathcal{A}_q|$. The expectation of $Z$ given a specific realization of hash functions, or equivalently, given $W$, is $\E(Z |W) = W$. 
\end{restatable}

As a consequence, it is immediately clear that $\E(W) = |\mathcal{A}_q|$. It is important to understand the implications of this lemma. In particular, the expression for $\E(Z|W)$ says that in a specific realization of a choice of hash functions (or a set of tables), the estimator $Z$ is biased if $W \neq |\mathcal{A}_q|$. Therefore $K$ is essential for helping concentrate the realized value of $W$ around its mean. 

\begin{proof}
We sample each $x \in \mathcal{A}_q$ with probability $\frac{\sum_{k=1}^K \mathds{1} (x\in \buckets(\mathcal{I})) }{\sum_{k=1}^K \countsvectorquery(\mathcal{I})}$. Given $W = \sum_{x \in \mathcal{A}_q} \frac{\sum_{k=1}^K \mathds{1} (x\in \buckets(\mathcal{I})) }{K \cdot p(x)}$, we have:
\begin{align*}
\E(Z | W) &= \sum_{x \in \mathcal{A}_q} \frac{\sum^K_{k=1}\countsvectorquery({\mathcal{I}})}{K \cdot p(x)} \cdot \frac{\sum_{k=1}^K \mathds{1} (x\in \buckets(\mathcal{I})) }{\sum_{k=1}^K \countsvectorquery(\mathcal{I})} \Bigg| W\\
& = \sum_{x \in \mathcal{A}_q} \frac{\sum_{k=1}^K \mathds{1} (x\in \buckets(\mathcal{I}))}{K \cdot p(x)} \Bigg| W\\
& = W
\end{align*}
Now, 
\begin{align*}
\E(W) &= \frac{1}{K} \sum\limits_{k=1}^K \sum_{x \in \mathcal{A}_q} \frac{\E(\mathds{1}(x \in \buckets(\mathcal{I})))}{p(x)} \\
& = \frac{1}{K} \sum\limits_{k=1}^K \sum_{x \in \mathcal{A}_q} \frac{p(x)}{p(x)} =  |\mathcal{A}_q|
\end{align*}
Then clearly, $\E(Z | W) = W$ and $\E(Z) = \E(\E(Z | W)) = |\mathcal{A}_q|$. 
\end{proof}

Since in expectation, our estimator $Z$ gives $W$, we want to understand how many tables $K$ are required to ensure that $W$ concentrates around its mean, $|\mathcal{A}_q|$. This is related to the variance of $W$.

We also introduce a new quantity $p(x,y) = \P(d_{qx} \in \mathcal{I} \cap d_{qy} \in \mathcal{I}| \theta_{qx}, \theta_{qy})$, the collision probability that $x$ and $y$ both land in buckets that are hamming distance $\mathcal{I}$ from $q$ over the random choice of hash functions. 

\begin{restatable}[Variance of $W$]{lemma}{variancew}
\label{lem:variance_w}
$\sigma^2(W) = \frac{1}{K} \sum\limits_{x,y \in \mathcal{A}_q} \left(\frac{p(x,y)}{p(x)p(y)} -1 \right)$~.
\end{restatable}

\begin{proof}
We want to compute: 
\begin{align*}
\E[W^2] &= \frac{1}{K^2} \sum_{k=1}^K \sum_{{x \in \mathcal{A}_q} \atop {y \in \mathcal{A}_q}} \frac{ \E(\mathds{1} (x\in \buckets(\mathcal{I}), y \in \buckets(\mathcal{I}))) }{p(x)p(y)} + \frac{1}{K^2} \sum_{k=1}^K \sum_{l \not= k} \sum_{{x \in \mathcal{A}_q} \atop {y \in \mathcal{A}_q}} \frac{\E[\mathds{1} (x\in \buckets(\mathcal{I})) \mathds{1} (y\in \mathcal{B}^l_q(\mathcal{I}))]}{p(x)p(y)}\\
& =\frac{1}{K} \sum_{x,y \in \mathcal{A}_q} \frac{p(x,y)}{p(x)p(y)} + \frac{1}{K^2}\sum_{k=1}^K \sum_{l \not= k} \sum_{{x \in \mathcal{A}_q} \atop {y \in \mathcal{A}_q}} \frac{p(x)p(y)}{p(x)p(y)}\\
&= \frac{1}{K} \sum_{x,y \in \mathcal{A}_q} \frac{p(x,y)}{p(x)p(y)} + \left(1-\frac{1}{K}\right) |\mathcal{A}_q|^2\\
\end{align*}
Since $\sigma^2(W) = \E[W^2] - (\E[W])^2$, we appeal to Lemma \ref{lem:expectation_z} to conclude: 
\[
\sigma^2(W) = \frac{1}{K} \sum\limits_{x,y \in \mathcal{A}_q} \left(\frac{p(x,y)}{p(x)p(y)} -1 \right) ~. 
\]
\end{proof}

We want to put these pieces together to make a statement about the number of tables $K$ we should create to guarantee low inherent bias in \firstEstimator. We use Chebyshev's Inequality to bound $W$'s deviation from its mean as a function of $K$ with a constant failure probability $\frac{1}{8}$. For simplicity, we fix a constant failure probability that we will boost later by average over several sets of estimators. This analysis is without loss of generality, as the bounds can be adjusted for any desired failure probability $\delta$. We will use this piece again when we do the full complete analysis for \firstEstimator. 

\begin{restatable}[Bound on Number of Tables]{lemma}{kbound}
\label{lem:K_bound}
It suffices to make $K \geq \frac{8}{\eps^2 \min\limits_{x \in \mathcal{A}_q} p(x)}$ tables to guarantee that $W$ is within $\eps$ of $|\mathcal{A}_q|$ (relatively) with probability $\frac{7}{8}$. 
\end{restatable}

\begin{proof}
Chebyshev's inequality states that
$\P(|W - |\mathcal{A}_q| | \geq \eps  |\mathcal{A}_q|) \leq \frac{\sigma^2(W)}{\eps^2 |\mathcal{A}_q|^2}$~.
Therefore, to achieve a constant failure probability $\delta = \frac{1}{8}$, it suffices to create enough tables so that 
\[
\sigma^2(W) = \frac{1}{K} \sum\limits_{x,y \in \mathcal{A}_q} \left(\frac{p(x,y)}{p(x)p(y)} -1 \right)  \leq \frac{\eps^2 |\mathcal{A}_q|^2}{8}~.
\]
Hence $K$ needs to be large enough so that
\[
K \geq \frac{8\sum\limits_{x,y \in \mathcal{A}_q} \left(\frac{p(x,y)}{p(x)p(y)} -1 \right) }{\eps^2 |\mathcal{A}_q|^2}~.
\]
Since $p(x,y) \leq \min\{p(x), p(y)\}$, we see that it is sufficient for $K$ to satisfy
\[
K \geq \frac{8 |\mathcal{A}_q|^2 \left(\frac{1}{\min_{x \in \mathcal{A}_q}p(x)} -1 \right) }{\eps^2 |\mathcal{A}_q|^2}~.
\]
Therefore we conclude with the following bound on $K$:  
\begin{equation}\label{K_bound_final}
K \geq \frac{8}{\eps^2 \min\limits_{x \in \mathcal{A}_q} p(x)}~.
\end{equation}
\end{proof}

We emphasize that the joint probability $p(x,y) \leq \min\{p(x), p(y)\}$ is a very loose worst-case bound assuming high correlation between data points. The final bound for $K$, Equation~\eqref{K_bound_final}, is also a worst-case bound in the sense that it is possible that a very minuscule fraction of $x \in \mathcal{A}_q$ have small values for $p(x)$. In the experimental section of the paper, we do an empirical analysis of the inherent bias for different values of $K$ and demonstrate that for real datasets the number of tables needed can be far fewer than what is theoretically required in the worst case scenario. We also give a finer-grained analysis for $p(x,y)$ in Appendix \ref{appendix_D}. 

\subsection{Counts Vector}
Query $q$ maps to a bucket $i^k$ for each table $k = 1, 2 \ldots K$. The preprocessing step produces an average counts vector corresponding to bucket $i^k$, denoted $\countsvectorquery$, where $\countsvectorquery (i)$ is the count of the total number of items in buckets that are at hamming distance $i=0, 1, \ldots t$ away from the hash address for $q$ in table $k$. For the hamming distances of interest $\mathcal{I}$, we let $\countsvectorquery(\mathcal{I}) = \sum_{d \in \mathcal{I}} \countsvectorquery(d)$. 

$\countsvectorquery({\mathcal{I}})$ is an integral part of our weighted importance sampling scheme. In Appendix \ref{appendix_A}, we show how to compute these vectors efficiently.

\begin{restatable}[$\aggregatecounts$]{theorem}{aggregatecountstheorem}
\label{thm:aggregate_counts_theorem}
Given a set of $K$ hash tables, each with $2^t$ hash buckets with addresses in $\{0,1\}^t$, $\aggregatecounts$ (Algorithm ~\ref{alg:aggregate_counts}) computes, for each hash address $i$, the number of elements in buckets that are hamming distance $0, 1, \ldots t$ away from $i$, in each of the $K$ tables, in time $O(Kt^2 2^t)$. 
\end{restatable}

Note that the $t$ in our hashing scheme is the length of the hash sequence; as a general rule of thumb, for bit-wise hash functions, implementers choose $t \approx \log(n)$, so as to average out to one element per hash bucket. Therefore, the preprocessing runtime of a reasonable hashing implementation for $\aggregatecounts$ (Algorithm ~\ref{alg:aggregate_counts}) is approximately $O(n K \log^2(n))$. 

The key benefit of $\aggregatecounts$ is that it computes via a message-passing or dynamic programming strategy that is much more efficient than a naive brute-force approach that would take time $O(K2^{2t})$, or $O(Kn^2)$ if $t \approx \log(n)$. 

\subsection{Sampler}

We create a sampler that, given a hash address $i^k$ for each table, and a set of hamming distances $\mathcal{I}$ that we want to sample from, generates a sample uniformly at random from the union of elements that were hashed to hamming distance $\mathcal{I}$ across the $K$ tables. For an implementation and analysis, please consult Appendix \ref{appendix_B}. 
\begin{restatable}[Sampler]{theorem}{sampler}
\label{thm:sampler}
Given a set of $K$ hash tables, each with $2^t$ hash buckets with addresses in $\{0,1\}^t$, a sampling scheme consisting of a data structure and a sampling algorithm can generate a sample uniformly at random from any fixed hash table $k$, an element at hamming distance $d$ to hash address $i$. The data structure is a counts matrix that can be precomputed in preprocessing time $O(K t^3 2^t)$, and the sampling algorithm $\hammingdistancesampler$ (Algorithm \ref{alg:hamming_distance_sampler}) generates a sample in time $O(t)$. 
\end{restatable}

Again, if we follow $t \approx \log(n)$, the preprocessing time comes out to roughly $O(nK \log^3(n))$. Also we expect the $O(t)$ online sample generation cost to be negligible compared to, say, the inner product computation cost for $q \cdot x$, which our method and all competing methods use. Now that we are able to generate samples, we describe and analyze the importance sampling scheme, \firstEstimator, in the next section. 

\section{\firstEstimator} \label{sec:sampling}

We now analyze \firstEstimator. Recall that \firstEstimator\space works in the following way. Given query $q$, we generate the hash for $q$ in each of our $K$ tables, by solving for $i^k = g_k(q)$ for $k=1, \ldots K$. Given the hash for $q$ in each of our $K$ tables and the set of hamming distances $\mathcal{I}$ that we want to sample from, we invoke our sampler to generate a sample from across the $K$ tables. 

Given this sample, $x$, we compute the angular distance $\theta_{qx} = \cos^{-1} (q \cdot x)$. Let $p(x) = \P(d_{qx} \in \mathcal{I} | \theta_{qx})$, the collision probability that $x$ lands in a bucket that is hamming distance $\mathcal{I}$ from $q$ over the random choice of hash functions; $p(x)$ is an endogenous property of an LSH function. 

We score each sample as in Equation~\eqref{eq:score_function}. We take $S$ samples and construct $Z_1, Z_2, \ldots Z_S$. We report $\frac{\sum_{i=1}^S Z_i}{S}$ as our estimate for $|\mathcal{A}_q|$. As an immediate consequence of Lemma $\ref{lem:expectation_z}$, it is clear that 
\[
\E\left[\frac{\sum_{i=1}^S Z_i}{S}\right] =|\mathcal{A}_q| ~.
\]
Now we analyze the variance of \firstEstimator: 

\begin{restatable}[Variance of Estimator]{lemma}{varianceestimator}
\label{eq:variance_estimator_analysis}
\[
\E\left[\left(\frac{\sum_{i=1}^S Z_i}{S} - |\mathcal{A}_q|\right)^2\right] \leq \frac{\E[Z^2]}{S} + \sigma^2(W)~.
\]
\end{restatable}

This decomposition of the variance into the two terms indicates that the variance is coming from two sources. The first source is the variance of the samples, $\frac{\E[Z^2]}{S}$. If we don't take enough samples, we do not get a good estimate. The second source is the variance from the random variable $W$, $\sigma^2(W)$, which corresponds to the contents in the tables. As we have shown, it is crucial to create enough tables so that $W$ is concentrated around its expectation, $|\mathcal{A}_q|$. Therefore, this second source of variance of \firstEstimator\space comes from the variance of the hash functions that underlie table creation and composition. 

\begin{proof}
The variance can be expressed as: 
\begin{align*}
\E\left[\left(\frac{\sum_{i=1}^S Z_i}{S} - |\mathcal{A}_q|\right)^2\right] &= 
\E\left[\left(\frac{\sum_{i=1}^S (Z_i-W)}{S} +(W-|\mathcal{A}_q|)\right)^2\right]\notag\\
&= \E\left[\frac{\sum_{i=1}^S (Z_i-W)^2}{S^2} +(W-|\mathcal{A}_q|)^2\right]\notag\\
&= \E\left[\frac{(Z-W)^2}{S} + (W-|\mathcal{A}_q|)^2\right]\notag\\
&= \frac{\E[Z^2]-\E[W^2]}{S} + \E[W^2]-|\mathcal{A}_q|^2 \notag\\
&\leq \frac{\E[Z^2]}{S} + \E[W^2]-|\mathcal{A}_q|^2 \notag\\
&= \frac{\E[Z^2]}{S} + \sigma^2(W) \label{eq:variance_estimator_analysis}
\end{align*}
\end{proof}

The $\sigma^2(W)$ term has already been analyzed in Section \ref{sec:hash_tables}, see Lemma \ref{lem:variance_w}. Now we analyze the second moment of $Z$. 

\begin{restatable}[Variance of $Z$]{lemma}{variancez}
\label{eq:z_squared}
\[
\E[Z^2] = \sum_{x \in \mathcal{A}_q}\sum_{y \in \mathcal{D}} \left[ \frac{p(x,y)}{K \cdot p(x)^2} + \left(1-\frac{1}{K}\right) \frac{p(y)}{p(x)} \right]~.
\]
\end{restatable}

\begin{proof}
To analyze the second moment of $Z$, as with our first moment analysis of $Z$, we first condition on fixing the hash tables, so given $g_1, \ldots g_K$, we know which elements of interest in $\mathcal{A}_q$ end up in our hamming distance set of interest $\mathcal{I}$.  

\begin{align*}
\E[Z^2 | g_1, \ldots g_K] &= \frac{1}{K^2} \left(\sum^K_{k=1}\countsvectorquery({\mathcal{I}})\right)\left(\sum^K_{k=1} \sum_{x \in \mathcal{A}_q \cap \buckets(\mathcal{I})} \frac{1}{p(x)^2}\right) \\
& = \frac{1}{K^2} \left[ \sum^K_{k=1} \sum_{{x \in \mathcal{A}_q \cap \buckets} \atop {y \in \buckets}} \frac{1}{p(x)^2} + \sum^K_{k=1}  \sum_{l \neq k} \sum_{{x \in \mathcal{A}_q \cap \buckets} \atop {y \in \mathcal{B}^l_q}} \frac{1}{p(x)^2}\right]
\end{align*}
Now using the fact that $\E[Z^2] = \E[\E[Z^2 | g_1, \ldots g_K]]$, we have: 
\begin{align}
\E[Z^2] &= \frac{1}{K^2} \E \left[ \sum^K_{k=1} \sum_{{x \in \mathcal{A}_q} \atop {y \in \mathcal{D}}} \frac{\mathds{1}(x \in \buckets(\mathcal{I}))\mathds{1}(y \in \buckets(\mathcal{I}))}{p(x)^2}\right] + \frac{1}{K^2} \E \left[ \sum^K_{k=1} \sum_{l \neq k} \sum_{{x \in \mathcal{A}_q} \atop {y \in \mathcal{D}}} \frac{\mathds{1}(x \in \buckets(\mathcal{I}))\mathds{1}(y \in \mathcal{B}^l_q(\mathcal{I}))}{p(x)^2}\right] \notag\\
&= \frac{1}{K} \sum_{{x \in \mathcal{A}_q} \atop {y \in \mathcal{D}}} \frac{p(x,y)}{p(x)^2} + \left(1 - \frac{1}{K}\right) \sum_{{x \in \mathcal{A}_q} \atop {y \in \mathcal{D}}} \frac{p(x)p(y)}{p(x)^2} \notag\\
&= \sum_{x \in \mathcal{A}_q}\sum_{y \in \mathcal{D}} \left[ \frac{p(x,y)}{K \cdot p(x)^2} + \left(1-\frac{1}{K}\right) \frac{p(y)}{p(x)} \right]
\end{align}
\end{proof}

Now that we have all the components, we are ready to put together the final sample and storage complexities for our estimator. We want a final estimate that concentrates with at most $\epsilon$ error around its mean, $|\mathcal{A}_q|$ with probability $1-\delta$. To do this, we make several sets $1, 2, \ldots M$ of our estimator (one estimator consists of a set of $K$ tables and $S$ samples). We choose $K$ and $S$ so that the failure probability of our estimator is a constant, say $\frac{1}{4}$. Each estimator produces an estimate, call it $E_m$, for $m \in \{1, \ldots M\}$. We report our final estimate as the median of these estimates. This is the classic Median-of-Means technique. 

Let $F_m$ be the indicator variable indicating if the estimator $E_m$ fails to concentrate. Clearly $\E(F_m) \leq \frac{1}{4}$. Moreover, $\E(F = \sum^M_{m=1} F_m) \leq \frac{M}{4}$. The probability that the median estimate is bad, $\P(\text{median of } E_m \text{fails}) \leq \P(\text{half of } E_m \text{ fails}) = \P(F \geq \frac{M}{2})$. By a simple Chernoff bound, we see that: 
$\P(F \geq \frac{M}{2}) \leq e^{-(2 \ln 2 -1) \frac{M}{4}} \leq e^{\frac{-M}{11}}$. So to satisfy a desired failure probability $\delta$, it suffices to have $e^{\frac{-M}{11}} \leq \delta$, therefore $M \in O(\log(\frac{1}{\delta}))$. 

In the rest of the section, we establish bounds on $K$ and $S$ so that one estimator fails with probability at most $\frac{1}{4}$. 
We appeal again to Chebyshev's Inequality: 
\[
\P\left(\Bigg| \frac{\sum_{i=1}^S Z_i}{S} - |\mathcal{A}_q| \Bigg| \geq \eps |\mathcal{A}_q| \right) \leq \frac{\sigma^2(\frac{\sum_{i=1}^S Z_i}{S})}{\eps^2 |\mathcal{A}_q|^2}~.
\]

In Lemma~\ref{eq:variance_estimator_analysis}, we analyze the variance of \firstEstimator, and show that $\sigma^2(\frac{\sum_{i=1}^S Z_i}{S}) \leq \frac{\E[Z^2]}{S} + \sigma^2(W)$. Therefore, in order so that the failure probability is less than $\frac{1}{4}$, it suffices to have $\sigma^2(\frac{\sum_{i=1}^S Z_i}{S}) \leq \frac{\eps^2 |\mathcal{A}_q|^2}{4}$, which can be obtained by letting $\frac{\E[Z^2]}{S} \leq \frac{\eps^2 |\mathcal{A}_q|^2}{8}$ and $\sigma^2(W) \leq \frac{\eps^2 |\mathcal{A}_q|^2}{8}$. 

Focusing on the $\sigma^2(W)$ term, which depends on the number of tables $K$ created, we show in Lemma~\ref{lem:K_bound} from Section \ref{sec:hash_tables} that it suffices to take $K \geq \frac{8}{\eps^2 \min\limits_{x \in \mathcal{A}_q} p(x)}$. 

Now that we have our table complexity, we can analyze our sampling complexity $S$ to bound $\frac{\E[Z^2]}{S}$.

\begin{lemma}
\label{final_sample}
Suppose $K \geq \frac{8}{\eps^2 \min\limits_{x \in \mathcal{A}_q} p(x)}$. Then $S \in O\left(\frac{\E(\countsvectornotablequery({\mathcal{I}}))}{\eps^2|\mathcal{A}_q| \cdot \min\limits_{x \in \mathcal{A}_q}p(x)}\right)$ suffices to achieve $\frac{\E[Z^2]}{S} \leq \frac{\eps^2 |\mathcal{A}_q|^2}{8}$. 
\end{lemma}

\begin{proof}
By Lemma~\ref{eq:z_squared} we have: 
$$\frac{\E[Z^2]}{S} = \frac{1}{S} \sum_{x \in \mathcal{A}_q}\sum_{y \in \mathcal{D}} \left[ \frac{p(x,y)}{K \cdot p(x)^2} + \left(1-\frac{1}{K}\right) \frac{p(y)}{p(x)} \right]$$
Substituting for $K \geq \frac{8}{\eps^2 \min\limits_{x \in \mathcal{A}_q} p(x)}$ gives: 
\begin{align*}
\frac{\E[Z^2]}{S} & \leq \frac{1}{S} \sum_{x \in \mathcal{A}_q}\sum_{y \in \mathcal{D}} \left[ \frac{\eps^2 p(x,y)\min\limits_{x \in \mathcal{A}_q} p(x)}{8p(x)^2} + \frac{p(y)}{p(x)} \right] \\
& \leq \frac{1}{S} \sum_{x \in \mathcal{A}_q}\sum_{y \in \mathcal{D}} \left[ \frac{\eps^2 p(x,y)}{8p(x)} + \frac{p(y)}{p(x)} \right] \\
& \leq \frac{1}{S}\sum_{x \in \mathcal{A}_q}\sum_{y \in \mathcal{D}} \left[ (1 + \eps^2) \frac{ p(y)}{p(x)} \right]
\end{align*}
In order to guarantee $\frac{\E[Z^2]}{S} \leq \frac{\eps^2 |\mathcal{A}_q|^2}{8}$, we need: 
\begin{align*}
S &\geq \frac{\sum_{x \in \mathcal{A}_q}\sum_{y \in \mathcal{D}} \left[ (1 + \eps^2) \frac{ p(y)}{p(x)} \right]}{\eps^2 |\mathcal{A}_q|^2} \\
& = (1 + \eps^2) \frac{\sum_{x \in \mathcal{A}_q} \frac{1}{p(x)} \sum_{y \in \mathcal{D}} p(y)}{\eps^2 |\mathcal{A}_q|^2} \\ 
& = (1 + \frac{1}{\eps^2}) \frac{\sum_{x \in \mathcal{A}_q} \frac{1}{p(x)} \E(\countsvectornotablequery({\mathcal{I}}))}{|\mathcal{A}_q|^2} 
\end{align*}
Therefore, we conclude that 
\[
S = O\left(\frac{\E(\countsvectornotablequery({\mathcal{I}}))}{\eps^2|\mathcal{A}_q| \cdot \min\limits_{x \in \mathcal{A}_q}p(x)}\right)
\]
is sufficient. 
\end{proof}

Putting together Lemmas \ref{lem:K_bound} and \ref{final_sample} with the median of means strategy yields our main result, Theorem~\ref{thm:mainresult}. 

We discuss the results of our experiments on real datasets using \firstEstimator\space in Section \ref{sec:experiments}. In the next section, we analyze our second estimator, \secondEstimator.

\section{\secondEstimator} \label{sec:multi-probe_importance}
We now describe another estimation scheme, \secondEstimator, that combines ideas from \cite{lv2007multi} with \firstEstimator\space that we introduce in this paper. We show in the experiments section that this estimator outperforms our previous estimator and all benchmarks for queries with very small neighborhoods. For larger neighborhoods, \firstEstimator\space seems to be more competitive. 

\subsection{Approach Overview}
The idea of \secondEstimator\space is to use the same importance weighing technique used in the \firstEstimator, however, instead of sampling from hamming distances $\mathcal{I}$, we search promising buckets according to the query-directed probing idea introduced in \cite{lv2007multi}. \secondEstimator\space inspects the elements in the chosen promising buckets, it does not sample, this is one key difference between \secondEstimator\space and \firstEstimator. 

The query-directed probing idea relies heavily on computing fine-grained collision probabilities for each bucket in the table, given the value of the query's projection onto each random hyperplane underlying the hash table. We provide a calculation for random hyperplane LSH in Appendix \ref{appendix_C} that we use to implement \secondEstimator.

Let $\bucketsMulti$ denote the set of hash buckets to multi-probe from in the $k$-th hash table. Let $\projectionsMulti$ denote values of the projection of the query onto the random hyperplanes for table $k$. We then score each element found as: 
\begin{equation}\label{eq:score_function_multiprobe}
Z = \begin{cases} 
      \frac{1}{\sum\limits_{k \in K} p^k_m(x)} & \text{ if } \theta_{qx} \in \mathcal{A} \\
     0 & \text{ otherwise. }
   \end{cases}
\end{equation}
where $p^k_m(x)$ is the probability that $x$ is hashed to the set of buckets $\bucketsMulti$, that is $p^k_m(x) = \P(x \in \bucketsMulti | \theta_{qx}, \projectionsMulti)$. Appendix \ref{appendix_C} gives the formula for computing this probability for random hyperplane LSH. 

For the $S$ elements that we inspect, we take $Z_1, Z_2, \ldots Z_S$ and report $\sum^S_{i=1} Z_i$ as the final estimate. Note that the same element can be found while inspecting buckets from multiple tables, that is fine, we score each and factor them into the final estimate, there is no need to remove duplicates in \secondEstimator.

One final note is that the resource of focus in \secondEstimator\space is bucket-based, so our analysis will be on what kind of criteria the buckets probed need to satisfy to guarantee a quality estimate. In this case, since multi-probe specifically is able to rank buckets according to their probability of containing an interesting element, our analysis will not give a specific number of buckets to probe, but rather give a quality criteria -- what the cumulative probabilities should be to produce a good estimate. This kind of analysis is a little more nebulous in that it does not immediately give space (table) and running time (sample) bounds. We show in the experimental section, Section \ref{sec:experiments} a range of different table and sample complexity configurations that implementers can consider. 

\subsection{Analysis}
Our estimator is essentially reporting a realization of the following random variable: 
\[
W_m = \sum\limits_{k=1}^K \sum_{x \in \mathcal{A}_q} \frac{\mathds{1}(x \in \bucketsMulti)}{\sum\limits_{k \in K} p^k_m(x)} ~,
\]
where $p^k_m(x) = \P(x \in \bucketsMulti | \theta_{qx}, \projectionsMulti)$. 

We first want to show that given the choice of buckets $\bucketsMulti$, $W_m$ is an unbiased estimator of $|{A}_q|$. 

\begin{lemma}[Expectation of $W_m$]
\label{lem:expectation_w_m}
$\E(W_m | \bucketsMulti, \projectionsMulti) = |{A}_q|~.$
\end{lemma}

We briefly remark here that it is important to condition on the choice of buckets, $\bucketsMulti$. In this scheme, the implementer chooses the target buckets, where the probability of the bucket containing an interesting element depends on the random hyperplanes. Therefore, while there is some randomness in which buckets are more or less promising, the algorithm explicitly needs the implementer to specify the buckets to probe, so the buckets are not random. The algorithm does not explicitly require that implementors choose the most promising buckets, though, intuitively and as we will see in the analysis, it is a good idea to do so for best performance. 

\begin{proof}
\begin{align*}
\E(W_m | \bucketsMulti, \projectionsMulti) &=\sum\limits_{k=1}^K \sum_{x \in \mathcal{A}_q} \frac{\E(\mathds{1}(x \in \bucketsMulti))}{\sum\limits_{k\in K}p^k_m(x)}\\
& = \sum_{x \in \mathcal{A}_q} \sum\limits_{k=1}^K\frac{p^k_m(x)}{\sum\limits_{k\in K}p^k_m(x)} =  |\mathcal{A}_q|
\end{align*}
\end{proof}

Now we want to understand how many buckets we should probe to ensure that our estimator concentrates around its mean, $|\mathcal{A}_q|$. This is related to the variance of $W_m$.

We also introduce a new quantity $p^k_m(x,y) = \P(x \in \bucketsMulti \cap x \in \bucketsMulti | \theta_{qx}, \theta_{qy}, \theta_{xy}, \projectionsMulti)$, the probability that $x$ and $y$ both land in $\bucketsMulti$. 

For simplicity of notation, we denote $p_m(x) = \sum\limits_{k\in K} p^k_m(x)$, and $p^k_m(x,y) = \sum\limits_{k\in K} p^k_m(x,y)$. 

\begin{restatable}[Variance of $W_m$]{lemma}{variancewm}
$\sigma^2(W_m | \bucketsMulti, \projectionsMulti) \leq\sum_{{x \in \mathcal{A}_q} \atop {y \in \mathcal{A}_q}}\left[ \frac{ p_m(x,y) }{p_m(x)p_m(y)}\right]$~.
\end{restatable}

\begin{proof}
We want to compute: 
\begin{align*}
\E[W^2_m | \bucketsMulti, \projectionsMulti] &= \sum_{k=1}^K \sum_{{x \in \mathcal{A}_q} \atop {y \in \mathcal{A}_q}} \frac{ \E(\mathds{1} (x\in \bucketsMulti, y \in \bucketsMulti)) }{p_m(x)p_m(y)} \\
&+ \sum_{k=1}^K \sum_{l \not= k} \sum_{{x \in \mathcal{A}_q} \atop {y \in \mathcal{A}_q}} \frac{\E[\mathds{1} (x\in \bucketsMulti) \mathds{1} (y\in \mathcal{B}^l_m(q))]}{p_m(x)p_m(y)}\\
& =\sum_{k=1}^K \sum_{{x \in \mathcal{A}_q} \atop {y \in \mathcal{A}_q}} \frac{ p^k_m(x,y) }{p_m(x)p_m(y)} + \sum_{k=1}^K \sum_{l \not= k} \sum_{{x \in \mathcal{A}_q} \atop {y \in \mathcal{A}_q}} \frac{p^k_m(x)p^l_m(y)}{p_m(x)p_m(y)}\\
& =\sum_{k=1}^K \sum_{{x \in \mathcal{A}_q} \atop {y \in \mathcal{A}_q}} \frac{ p^k_m(x,y) }{p_m(x)p_m(y)} + \sum_{{x \in \mathcal{A}_q} \atop {y \in \mathcal{A}_q}} \sum_{k=1}^K \sum_{l \not= k} \frac{p^k_m(x)p^l_m(y)}{p_m(x)p_m(y)}\\
&=\sum_{k=1}^K \sum_{{x \in \mathcal{A}_q} \atop {y \in \mathcal{A}_q}} \frac{ p^k_m(x,y) }{p_m(x)p_m(y)} + \sum_{{x \in \mathcal{A}_q} \atop {y \in \mathcal{A}_q}} \sum_{k=1}^K\frac{p^k_m(x)(p_m(y) - p^k_m(y))}{p_m(x)p_m(y)}\\
&=\sum_{k=1}^K \sum_{{x \in \mathcal{A}_q} \atop {y \in \mathcal{A}_q}} \frac{ p^k_m(x,y) }{p_m(x)p_m(y)} + \sum_{{x \in \mathcal{A}_q} \atop {y \in \mathcal{A}_q}} \sum_{k=1}^K\left[\frac{p^k_m(x)p_m(y)}{p_m(x)p_m(y)} - \frac{p^k_m(x)p^k_m(y)}{p_m(x)p_m(y)}\right]\\
&=\sum_{k=1}^K \sum_{{x \in \mathcal{A}_q} \atop {y \in \mathcal{A}_q}}\left[ \frac{ p^k_m(x,y) }{p_m(x)p_m(y)} - \frac{p^k_m(x)p^k_m(y)}{p_m(x)p_m(y)}\right]+ |\mathcal{A}_q|^2\\
&\leq\sum_{k=1}^K \sum_{{x \in \mathcal{A}_q} \atop {y \in \mathcal{A}_q}}\left[ \frac{ p^k_m(x,y) }{p_m(x)p_m(y)}\right]+ |\mathcal{A}_q|^2\\
\end{align*}
Since $\sigma^2(W_m | \bucketsMulti, \projectionsMulti) = \E[W^2_m | \bucketsMulti, \projectionsMulti] - (\E[W_m | \bucketsMulti, \projectionsMulti])^2$, we appeal to Lemma \ref{lem:expectation_w_m} to conclude: 
\[
\sigma^2(W_m | \bucketsMulti, \projectionsMulti) \leq\sum_{k=1}^K \sum_{{x \in \mathcal{A}_q} \atop {y \in \mathcal{A}_q}}\left[ \frac{ p^k_m(x,y) }{p_m(x)p_m(y)}\right] = \sum_{{x \in \mathcal{A}_q} \atop {y \in \mathcal{A}_q}}\left[ \frac{ p_m(x,y) }{p_m(x)p_m(y)}\right]\\  ~. 
\]
\end{proof}

The bound that we aim for in this estimator is a little bit different. Instead of establishing sufficient conditions for the number of tables we create, we should seek to establish conditions for the threshold cumulative probability of finding the interesting elements. This analysis does not immediately prescribe how many tables to make, but does so implicitly. Each table only has so many buckets with high probability of containing an interesting element, so after one table's buckets have been mostly exhausted, it is necessary to make another table. 

We use Chebyshev's Inequality to bound $W_m$'s deviation from its mean as a function of $p_m(x)$ with failure probability $\delta$. As before, we first fix $\delta = \frac{1}{8}$, which we will later reduce to $\delta$ using median of means. 

\begin{restatable}[Bound on Buckets Probed]{lemma}{kbound_wm}
It suffices to probe enough buckets such that $\min_{x \in \mathcal{A}_q} p_m(x) \geq \frac{8}{\eps^2}$ to guarantee that $W_m$ is within $\eps$ of $|\mathcal{A}_q|$ (relatively) with probability $\frac{7}{8}$. 
\end{restatable}

\begin{proof}
Chebyshev's inequality states that
$\P(|(W_m | \bucketsMulti, \projectionsMulti) - |\mathcal{A}_q| | \geq \eps  |\mathcal{A}_q|) \leq \frac{\sigma^2(W_m | \bucketsMulti, \projectionsMulti)}{\eps^2 |\mathcal{A}_q|^2}$~.
Therefore, to achieve a constant failure probability $\delta = \frac{1}{8}$, it suffices to create enough tables so that 
\[
\sigma^2(W_m | \bucketsMulti, \projectionsMulti) \leq \sum_{{x \in \mathcal{A}_q} \atop {y \in \mathcal{A}_q}}\left[ \frac{ p_m(x,y) }{p_m(x)p_m(y)}\right] \leq \frac{\eps^2 |\mathcal{A}_q|^2}{8}~.
\]
First we want to further bound $\sigma^2(W_m | \bucketsMulti, \projectionsMulti)$ so that it is easier to work with. We have: 
\[
\sum_{{x \in \mathcal{A}_q} \atop {y \in \mathcal{A}_q}}\left[ \frac{ p_m(x,y) }{p_m(x)p_m(y)}\right] \leq |\mathcal{A}_q|^2 \left( \frac{1}{\min_{x \in \mathcal{A}_q} p_m(x)}\right)
\]
Therefore we conclude with the following bound on $K$:  
$$\min_{x \in \mathcal{A}_q} p_m(x) \geq \frac{8}{\eps^2}~.$$
\end{proof}
The implication of the above bound says that for \secondEstimator, it is good to use the critical mass of buckets that will contain interesting elements for each different table.

After applying median of means, we achieve the following Main Result: 
\mainresultmulti*

\section{Experiments} \label{sec:experiments}

\begin{figure}[h]
\centering
\includegraphics[width=0.5\columnwidth]{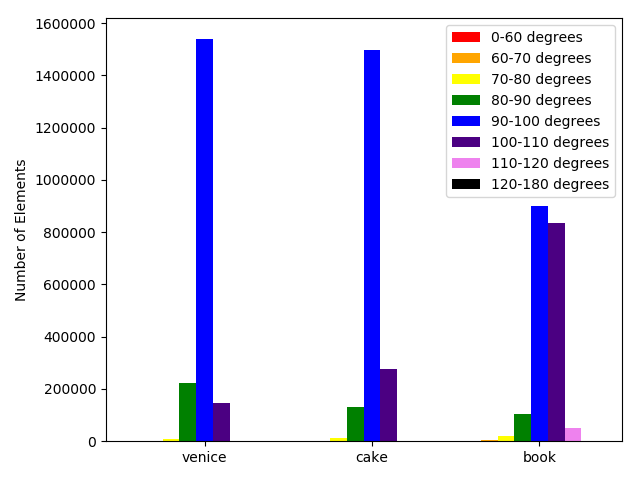}
\caption{The statistical profiles of angular distance between our 3 queries and the rest of the dataset $\mathcal{D}$. Notice that most (more than 99\%) embeddings in the dataset fall at about 60-120 degrees to the query. The leftmost red bins that represent the number of elements between 0-60 degrees to our 3 queries are barely noticeable.}
\label{fig:query-stats}
\end{figure}

We describe our experiments using the GLOVE dataset. We use the set of 1.9 million 300-dimensional word embedding vectors trained from Common Crawl, provided by \cite{pennington2014glove}. We normalize the embeddings, as is standard in many word embedding applications \cite{similar-word-embeddings}. We choose 3 query words with different neighborhood profiles: ``venice", ``cake", ``book". Venice has the smallest neighborhood, with 12 elements with angular distance less than 60 degrees, cake has a medium sized neighborhood with about 117 elements, book has the largest neighborhood with 424 elements. These are representative queries for the different types of regimes one may encounter. The histograms for these 3 queries are shown in Figure \ref{fig:query-stats}.

We also choose our angle range of interest, $\mathcal{A}$, to be 0-60 degrees. A search through our dataset gave ``florence", ``naples", ``rome" as representative elements that are 50-60 degrees from ``venice". Terms such as ``pie", ``dessert", ``cookie", and ``cheesecake" appear in the 40-50 degree annulus around ``cake", while terms such as ``shortcake", ``batter", ``tiramisu", ``bundt", and ``egg" appear in the 50-60 degree histogram. For ``book": ``author", ``ebook", ``paperback", ``reading", and  ``story" are in the 40-50 degree range while ``bestseller", ``literature", ``publisher", ``novel", ``edition", and ``chapter" are in the 50-60 degree range. This particular experiment shows that while elements in the 40-50 degree range are extremely related, words in the 50-60 degree range are also relevant, and so we fix $\mathcal{A}$ to be 0-60 degrees in all of our experiments. We also fix $t = 20$ in all of our experiments, since we have around 2 million embeddings in total and $20 \approx \log_2(2,000,000)$. 

As Table \ref{table:query_stats} illustrates, the biggest challenge for this estimation problem is the fact that the count of the number of elements within 0-60 degrees is dwarfed by the number of elements 60-120 degrees away from the queries. This issue makes locality sensitive techniques necessary for efficient search and retrieval in high dimensions. 

\begin{table}[h]
\caption{Statistics of Queries}
\label{table:query_stats}
\centering
\begin{tabular}{lccc}
\hline
Query & \# within 60 degrees & \% of population  \\
\hline
Venice   & 12    & .0006\\
Cake &   117  & .0061\\
Book & 424 & .0221\\
\hline
\end{tabular}
\end{table}

We implement 4 different techniques, \firstEstimator, \secondEstimator, and our two benchmark techniques, LSH-based estimator of \cite{Spring:2017aa} and Multi-Probe. We show that the estimators we develop improve upon the benchmarks for the each of the 3 regimes that our 3 fixed queries represent -- small, medium, and dense neighborhoods. We also compare \firstEstimator\space and \secondEstimator\space against each other, and show that \firstEstimator\space does better for dense neighborhoods, and \secondEstimator\space performs better for sparse neighborhoods. We first discuss the performance of \firstEstimator. 

\subsection{\firstEstimator}

\begin{figure}[h]
\centering
\begin{tabular}{ccc}
\includegraphics[width = .3\columnwidth]{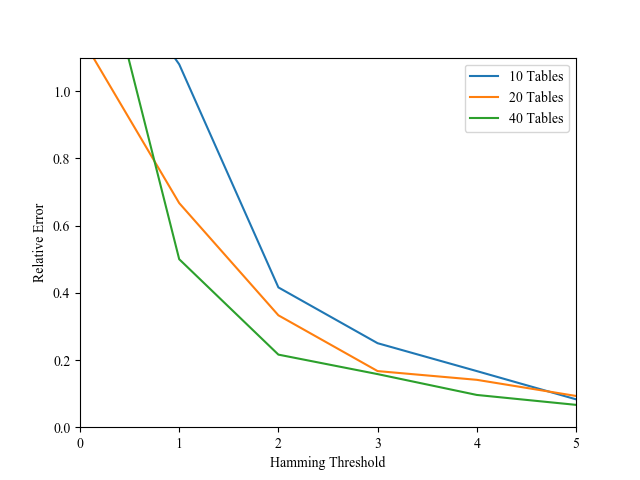} & \includegraphics[width = .3\columnwidth]{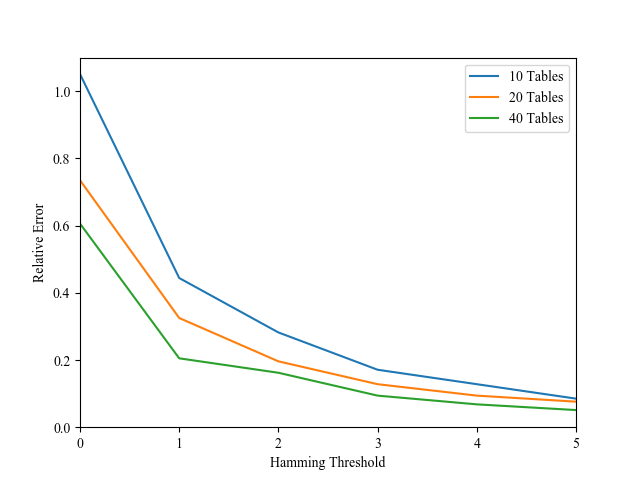} &
\includegraphics[width = .3\columnwidth]{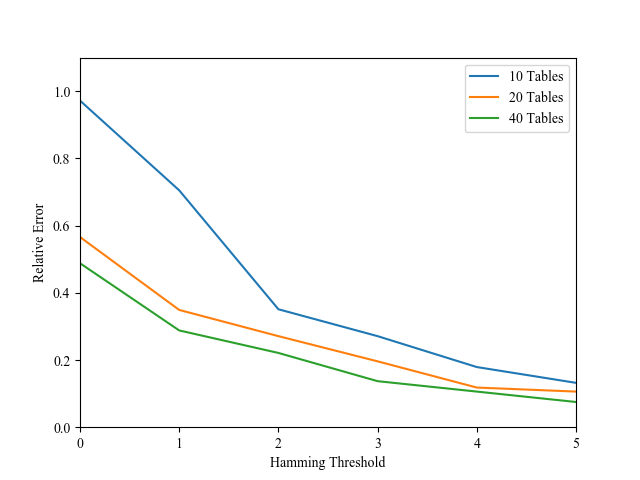}\\
\scriptsize{(a) venice} & \scriptsize{(b) cake} & \scriptsize{(c) book}
\end{tabular}
\caption{The empirical bias for different values of $K$ for queries ``venice", ``cake" and ``book". For each hamming threshold, the relative bias is averaged over 50 sets of $K$ tables, using random hyperplane hash as the LSH function.}
\label{fig:table-stats}
\end{figure}

As we have previously mentioned in section \ref{sec:hash_tables}, the number of tables $K$ theoretically required for (near) unbiased estimation in \firstEstimator\space relies on a worst-case variance bound; real-world data do not necessarily exhibit worst-case behavior. In our studies of our 3 queries in Figure \ref{fig:table-stats}, the inherent bias of \firstEstimator\space decreases as we increase the sampling hamming threshold. This is as expected -- using a larger range of hamming distances helps concentrate the count of the elements of interest $\mathcal{A}_q$ that fall into the specified range of hamming distances around the mean, which means that a smaller $K$ is required to achieve small bias. 

As Figure \ref{fig:table-stats} shows, the empirical bias of \firstEstimator\space at hamming threshold 5 is under 10\% for 20 hash tables, with very little improvement with 40 hash tables. This is consistent with our 3 queries, and demonstrates that for real datasets the number of tables needed can be far fewer than what is theoretically required in the worst case. 

\begin{figure}[h]
\centering
\centerline{\includegraphics[width=.5\columnwidth]{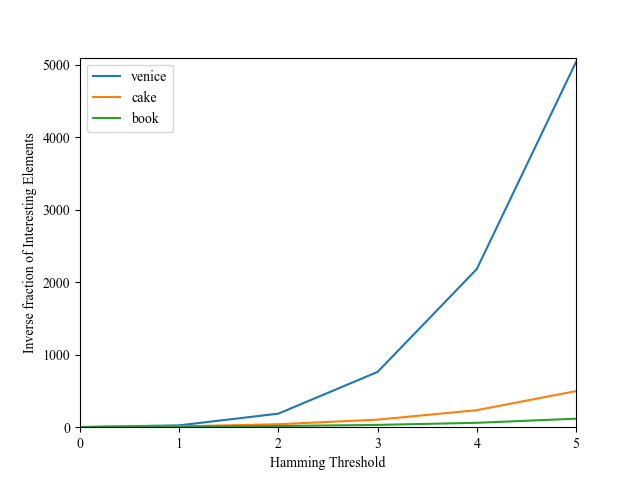}}
\caption{The inverse proportion of relative elements of interest in the overall sub-sampling pool for various hamming thresholds, averaged over 50 trials of sets of 20 hash tables.}
\label{fig:sample_complexity}
\end{figure}

In terms of sample complexity, it is worth noting that using higher thresholds typically requires more samples, as shown in Figure \ref{fig:sample_complexity}. However, higher thresholds typically lowers the inherent bias in the importance sampling scheme, as demonstrated in Figure \ref{fig:table-stats}. Implementers should consider this tradeoff in their algorithmic design choices.

\subsubsection{Benchmark 1: LSH-based estimator of \cite{Spring:2017aa}}
We compare \firstEstimator\space against the benchmark estimator introduced by \cite{Spring:2017aa}. Though their work originally intended to solve a different problem, their technique can solve for local density by adapting the weight function appropriately. The key differences between their work and \firstEstimator\space is that they only probe the 0 hamming distance bucket in each table, similar to the LSH application for nearest neighbor search, and instead of sampling, they simply enumerate the elements in the hamming distance 0 bucket for each table. 

\begin{figure}[h]
\centering
\begin{tabular}{cc}
\includegraphics[width = 0.4\columnwidth]{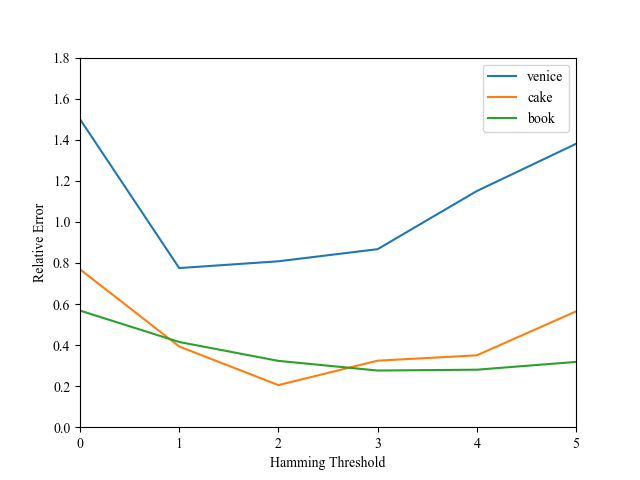} &
\includegraphics[width =  0.4\columnwidth]{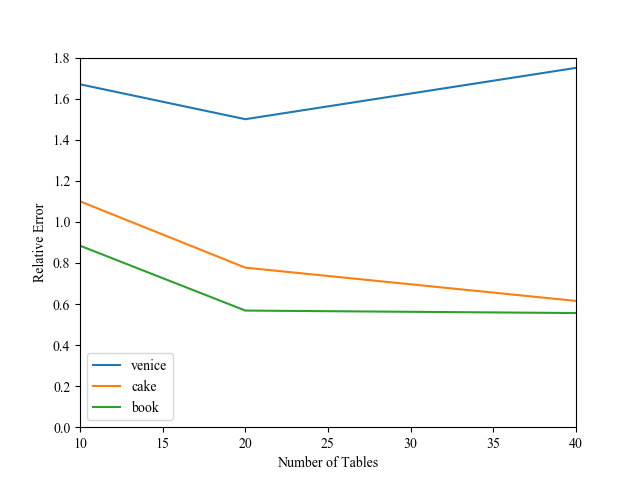}\\ \scriptsize{(a)
\firstEstimator\space fixing 20 Tables and 1000 samples} & \scriptsize{(b) \cite{Spring:2017aa} estimate fixing $\mathcal{I} = 0$}
\end{tabular}
\caption{Comparison of \firstEstimator\space against the benchmark LSH estimator adapted from ideas introduced in \cite{Spring:2017aa}.}
\label{fig:compare-benchmark}
\end{figure}

In Figure \ref{fig:compare-benchmark}, we compare \cite{Spring:2017aa}'s technique of enumerating and importance-weighting hamming distance 0 elements to our technique of importance sampling from different hamming thresholds. Our experiments use random hyperplane LSH and we report relative error averaged over 50 trials, where in each trial we generate a new set of $K$ tables. Panel (b) gives the results of experiments with \cite{Spring:2017aa}'s technique for the 3 queries, with different choices of $K$ (the number of tables). Our results show that even for $K = 40$ tables, the relative error of their technique can still be higher than 50\%, particularly for queries with small neighborhoods such as ``venice". For ``venice" the increase in table allocation from 20 to 40 made a very small difference to the overall estimation error. ``book" and ``cake" fared better at 40 tables, however, the error was still more than 50 \%. 

In comparison, \firstEstimator\space does very well using only 20 tables at the 2-3 hamming thresholds, estimating to within 20\% error. Panel (a) of Figure \ref{fig:compare-benchmark} shows that utilizing any hamming threshold greater than 0 gives superior estimation performance to staying only within the 0 hamming distance bucket. In this experiment, we fix our sampling budget to 1000 samples and the table budget to 20 tables. The hamming distance 0 error reported in this figure uses enumeration; all other hamming thresholds use the 1000 sampling budget. In our experiments for the 3 queries, one can expect about 50 points in total in the hamming distance 0 buckets across 20 tables. In this experiment, our technique uses 1000 samples vs 50 points, however, this (somewhat negligible in today's computing infrastructure) sample complexity trades off against a large improvement in accuracy.  

Finally, we note that panel (a) of Figure \ref{fig:compare-benchmark} shows that going to further hamming distances actually hurts the quality of the estimate. This phenomenon is related to the characteristics of the query and the sampling budget, because we actually dilute the proportion of interesting elements at higher thresholds.

\subsubsection{Benchmark 2: Multi-Probe}
We compare \firstEstimator\space against the technique introduced in \cite{lv2007multi}. Multi-probe was introduced to achieve smaller space complexity for approximate nearest neighbor search using locality sensitive hash functions. Its main premise is the idea of probing multiple buckets per hash table, since approximate near neighbors are very likely to also end up in buckets that are adjacent to the query bucket. The probing order for the buckets is determined based on the success probability of containing an interesting element. As an extension of \cite{lv2007multi}, we provide the collision probability calculations for multi-probe for random hyperplane projection in Appendix \ref{appendix_C}. 

\begin{figure}[h]
\centering
\begin{tabular}{cc}
\includegraphics[width = 0.4\columnwidth]{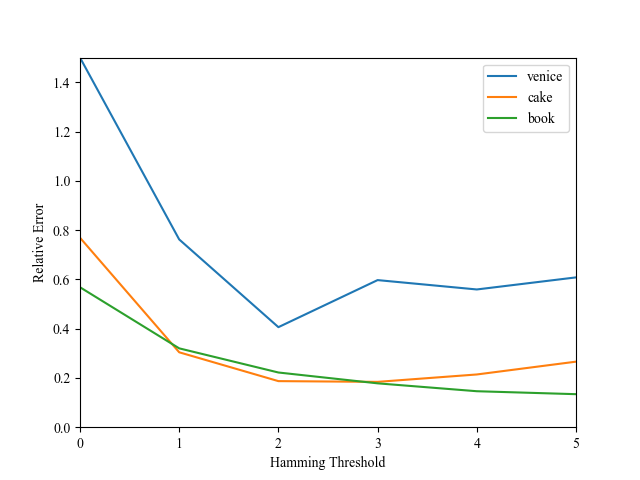} &
\includegraphics[width =  0.4\columnwidth]{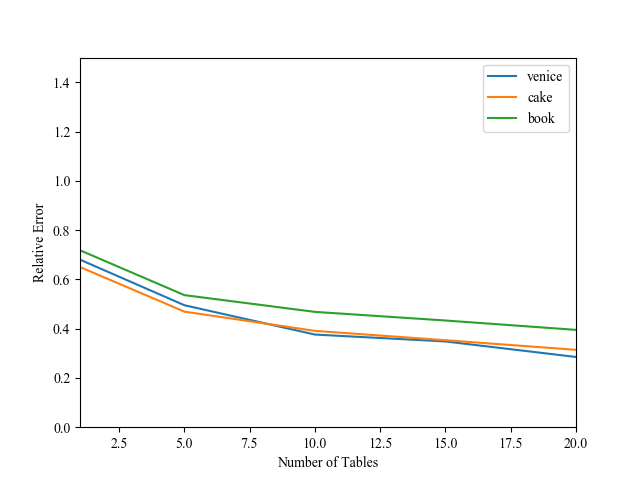}\\ \scriptsize{(a)
\firstEstimator\space Fixing 20 Tables and 5000 samples} & \scriptsize{(b) Multi-probe estimator fixing 5000 samples}
\end{tabular}
\caption{Comparison of \firstEstimator\space against multi-probe, fixing 5000 samples}
\label{fig:compare-benchmark-multiprobe-5000s}
\end{figure}

In Figures \ref{fig:compare-benchmark-multiprobe-1000s} and \ref{fig:compare-benchmark-multiprobe-5000s}, we compare \firstEstimator\space against multi-probe. The implementation of multi-probe, as presented in \cite{lv2007multi} and adapted for local density, would be to probe buckets according to their success probabilities, and report the number of interesting elements seen (after removing duplicates across multiple tables). In our experiments, we compare the effectiveness of drawing 1000 and 5000 samples in \firstEstimator\space, and restricting to retrieving 1000 and 5000 elements using multi-probe. As we show in Appendix \ref{appendix_C}, the bucket success probabilities for multi-probe depends on the target angle, and in our problem our target angle is actually a range of angles (0-60 degrees). We experiment with ranking buckets using fixed target angles 30, 45, and 60 degrees, and note negligible differences in performance. Therefore we show experimental results using fixed angle 45 degrees as the target for the multi-probe calculations.

As shown in Figures \ref{fig:compare-benchmark-multiprobe-1000s} and \ref{fig:compare-benchmark-multiprobe-5000s}, the performance of multi-probe suffers dramatically from the contamination of non-interesting elements into promising buckets close to the query bucket. Moreover, multi-probe only shows small improvements with increased number of tables. However, multi-probe performs better on queries with truly small neighborhoods, such as ``venice", with only 12 elements. For ``cake" and ``book", \firstEstimator\space performs better. 

\begin{figure}[h]
\centering
\begin{tabular}{cc}
\includegraphics[width = 0.4\columnwidth]{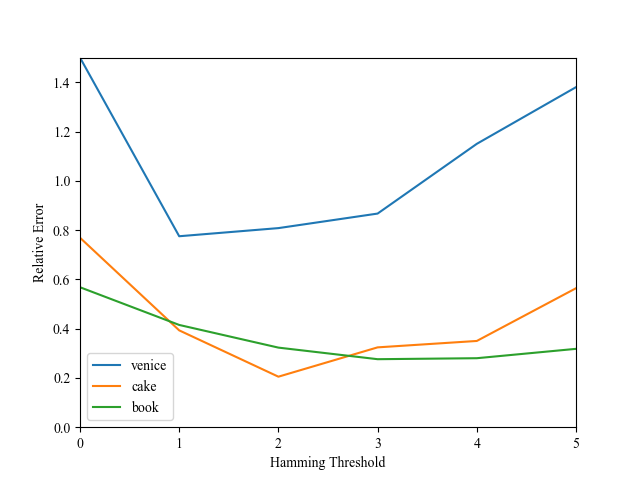} &
\includegraphics[width =  0.4\columnwidth]{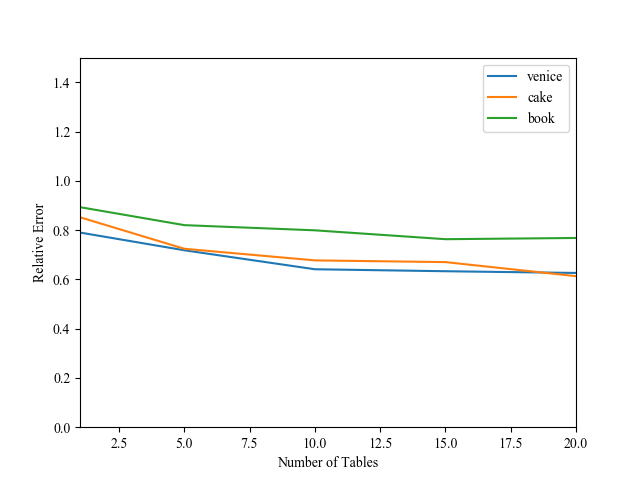}\\ \scriptsize{(a)
\firstEstimator\space Fixing 20 Tables and 1000 samples} & \scriptsize{(b) Multi-probe estimator fixing 1000 samples}
\end{tabular}
\caption{Comparison of \firstEstimator\space against multi-probe, fixing 1000 samples}
\label{fig:compare-benchmark-multiprobe-1000s}
\end{figure}

While \firstEstimator\space does not outperform multi-probe for very small queries, our multi-probe with importance weighting estimator, \secondEstimator, does. We discuss our experiments with \secondEstimator\space below. 

\subsection{\secondEstimator}
\begin{figure}[h]
\centering
\begin{tabular}{cc}
\includegraphics[width = 0.4\columnwidth]{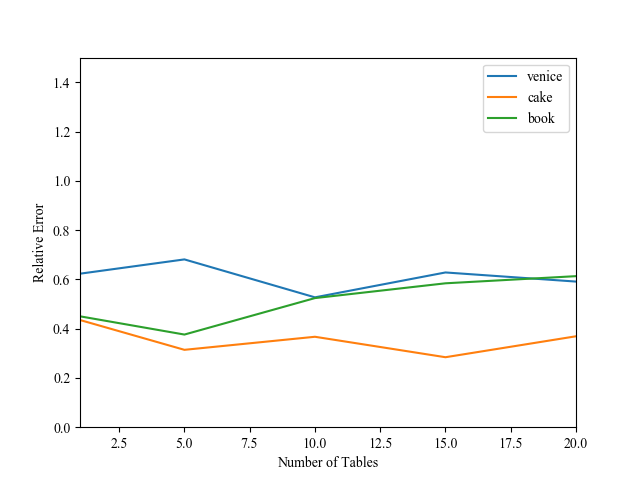} &
\includegraphics[width =  0.4\columnwidth]{Figures/multiprobe_45degrees_1000s}\\ \scriptsize{(a)
\secondEstimator\space Fixing 1000 samples} & \scriptsize{(b) Multi-probe estimator fixing 1000 samples}
\end{tabular}
\caption{Comparison of \secondEstimator\space against multi-probe, fixing 1000 samples}
\label{fig:compare-multi-probe-count-benchmark-multiprobe-1000s}
\end{figure}

\begin{figure}[h]
\centering
\begin{tabular}{cc}
\includegraphics[width = 0.4\columnwidth]{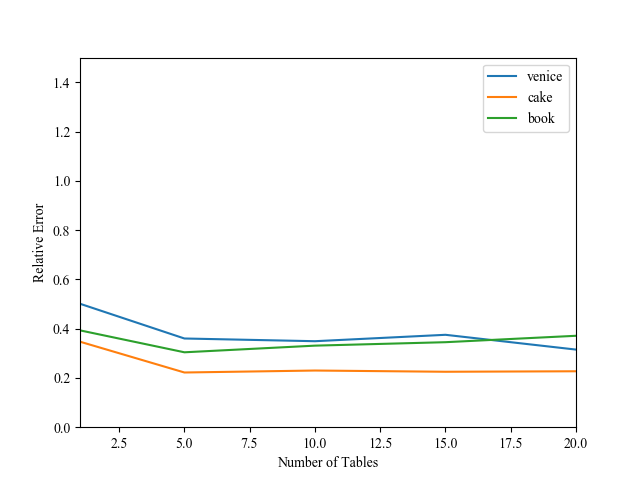} &
\includegraphics[width =  0.4\columnwidth]{Figures/multiprobe_45degrees_5000s}\\ \scriptsize{(a)
\secondEstimator\space Fixing 5000 samples} & \scriptsize{(b) Multi-probe estimator fixing 5000 samples}
\end{tabular}
\caption{Comparison of \secondEstimator\space against multi-probe, fixing 5000 samples}
\label{fig:compare-multi-probe-count-benchmark-multiprobe-5000s}
\end{figure}

We compare \secondEstimator\space against multi-probe. Our implementation of \secondEstimator\space and multi-probe are directly comparable -- the buckets chosen for probing are the buckets ranked most likely to contain an interesting element, where the target angle chosen for the bucket collision probability is 45 degrees. The only and key difference in the two estimators is the weighting scheme, \secondEstimator\space importance weighs each interesting element. Figure \ref{fig:compare-multi-probe-count-benchmark-multiprobe-1000s} shows that \secondEstimator\space outperforms multi-probe for all 3 queries, when both schemes are restricted to 1000 samples. When given a budget of 5000 samples, the results in figure \ref{fig:compare-multi-probe-count-benchmark-multiprobe-5000s} demonstrate that \secondEstimator\space outperforms multi-probe for the denser queries, ``cake" and ``book". For ``venice", \secondEstimator\space outperforms multi-probe up to 10 tables. When the sampling budget was shared between more than 10 tables, multi-probe does marginally better (1 or 2 percentage points). It is not clear if this margin is statistically significant. 

\subsection{Comparative Advantages of \firstEstimator\space and \secondEstimator}
\begin{figure}[h]
\centering
\begin{tabular}{cc}
\includegraphics[width = 0.4\columnwidth]{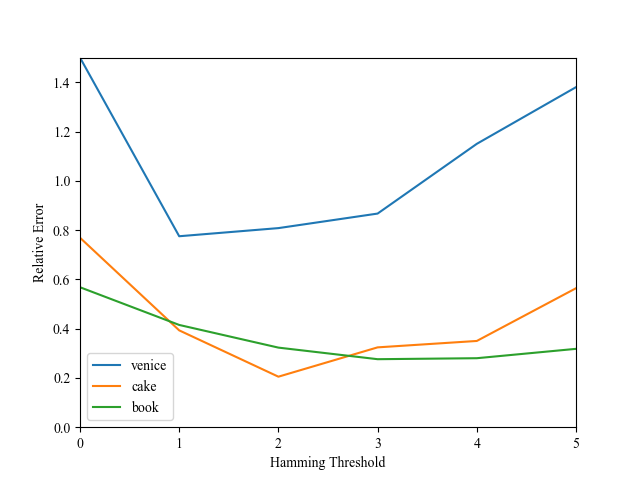} &
\includegraphics[width =  0.4\columnwidth]{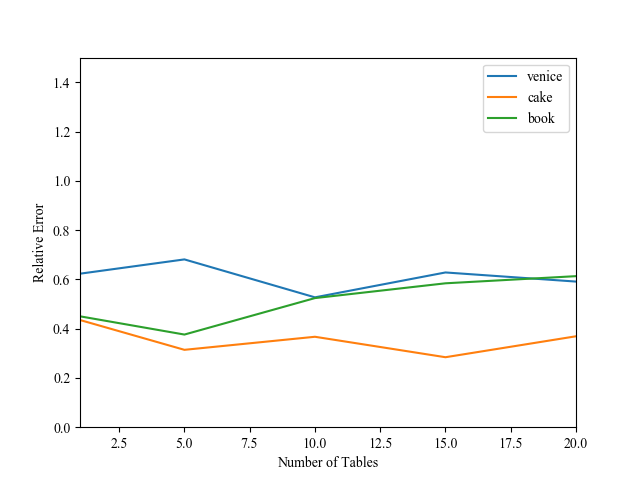}\\ \scriptsize{(a)
\firstEstimator\space Fixing 20 Tables and 1000 samples} & \scriptsize{(b) \secondEstimator\space Fixing 1000 samples}
\end{tabular}
\caption{Comparison of \firstEstimator\space and \secondEstimator, fixing 1000 samples}
\label{fig:compare-lsh-multi-probe-count-1000s}
\end{figure}

\begin{figure}[h]
\centering
\begin{tabular}{cc}
\includegraphics[width = 0.4\columnwidth]{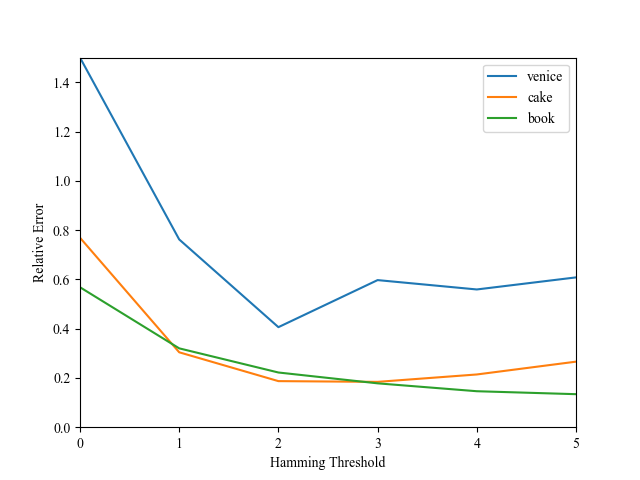} &
\includegraphics[width =  0.4\columnwidth]{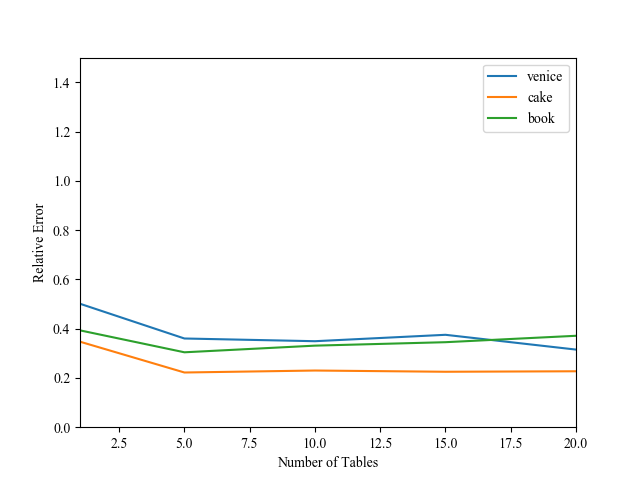}\\ \scriptsize{(a)
\firstEstimator\space Fixing 20 Tables and 5000 samples} & \scriptsize{(b) \secondEstimator\space Fixing 5000 sampless}
\end{tabular}
\caption{Comparison of \firstEstimator\space and \secondEstimator, fixing 5000 samples}
\label{fig:compare-lsh-multi-probe-count-5000s}
\end{figure}

We compare the two estimators that we develop in this paper, \firstEstimator\space and \secondEstimator\space side by side in Figures \ref{fig:compare-lsh-multi-probe-count-1000s} and \ref{fig:compare-lsh-multi-probe-count-5000s}. Figure \ref{fig:compare-lsh-multi-probe-count-1000s} shows the results of the two estimators, restricted to 1000 samples, while figure \ref{fig:compare-lsh-multi-probe-count-5000s} gives the results for 5000 samples. \secondEstimator\space outperforms \firstEstimator\space for ``venice," while \firstEstimator\space outperforms \secondEstimator\space for ``book." For ``cake," \firstEstimator\space outperforms \secondEstimator\space if one chooses a good hamming threshold to sample from for \firstEstimator, in this case, the better threshold is around 2 bits. 

\section{Closing Remarks}
We introduce a technique for estimating the number of points within a certain angular distance to a query for high dimensional vector datasets. We hope that this method will play an important role across many applications that use high dimensional vector embeddings as their data representation. We demonstrate that in the high dimensional setting, it is important to use techniques that are locality-sensitive and have dimension-free performance guarantees. We also employ a variety of variance reduction techniques, including using multiple hash tables, sampling from multiple buckets, and importance weighting our samples to further improve the performance of our estimator. We give provable bounds and demonstrate the effectiveness of our estimator in computational results. Further directions for research include better estimation schemes for a broader range of distance functions and types of datasets. 
\section*{Acknowledgements}
This work was initiated while the authors were visiting Laserlike, Inc. Xian Wu was supported by a Harold Thomas Hahn Jr. Fellowship from the Department of Management Science and Engineering at Stanford University. Moses Charikar was supported by NSF grant CCF-1617577 and a Simons Investigator Award. We thank Tatsunori Hashimoto and Virag Shah for helpful comments and feedback. 
\bibliography{distribution_estimation}
\bibliographystyle{alpha}
\appendix
\section{Computing the Average Counts Vector} \label{appendix_A}

We propose a message-passing algorithm, $\aggregatecounts$ (Algorithm ~\ref{alg:aggregate_counts}), which for a single table $k$ computes $\countsvector$ for all buckets $i$ in time $O(t^2 2^t)$. Repeating for each of the $K$ tables yields a total $O(Kt^2 2^t)$ runtime. $\aggregatecounts$ proceeds in $t$ rounds where in each round $r$, each bucket $i$ updates $\countsvector[r]$ by aggregating information passed from its neighbors in the set $\neighbors$. 

\aggregatecountstheorem*

\begin{algorithm}[ht]
\caption{Aggregate-Counts}
\label{alg:aggregate_counts}
\begin{algorithmic}[1]
\REQUIRE{Hash table with buckets $B$}
\FOR{round $r=0, r++, r \leq t$}
\FOR{hash address $i \in B$ }
\IF{r == 0}
\STATE $\countsvectornotable[r] = b_i$
\ELSIF{r ==1}
\STATE $\countsvectornotable[r] = \sum_{j \in \neighbors} \countsvectornotablej[r-1]$
\ELSE
\STATE $\countsvectornotable[r] = \frac{\sum_{j \in \neighbors} \countsvectornotablej[r-1] - (t-r+2)\countsvectornotable[r-2]}{r}$
\ENDIF
\ENDFOR
\ENDFOR
\STATE \textbf{Return} $\countsvectornotable$ for all $i \in B$
\end{algorithmic}
\end{algorithm}

\subsection{Analysis}
We first analyze the correctness of $\aggregatecounts$ (Algorithm ~\ref{alg:aggregate_counts}). Since the algorithm proceeds in rounds, we want to show that at the end of each round $r$, each $\countsvectornotable[0], \countsvectornotable[1], \ldots \countsvectornotable[r]$ correctly represents the number of elements in buckets of hamming distance $0, 1, \ldots r$ from address $i$, for all addresses $i \in B$. We use proof by induction on the number of rounds, $r$. 

\bigskip

\textbf{Base Case:} In round $r=0$, the update rule $\countsvectornotable[0] = b_i$ is clearly correct. For round $r=1$, the update rule $\countsvectornotable[1] = \sum_{j \in \neighbors} \countsvectornotablej[0]$ is also clearly correct. \\

\textbf{Inductive Hypothesis:} Assume that at end of round $r-1$, each $\countsvectornotable[0], \countsvectornotable[1], \ldots \countsvectornotable[r-1]$ correctly represents the number of elements in buckets of hamming distance $0, 1, \ldots r-1$ from address $i$, for all addresses $i \in B$. \\

\textbf{Inductive Step:} We want to show that at round $r$, $\countsvectornotable[r] = \frac{\sum_{j \in \neighbors} \countsvectornotablej[r-1] - (t-r+2)\countsvectornotable[r-2]}{r}$ correctly counts the number of elements in buckets at hamming distance $r$ from bucket $i$.  

Without loss of generality, we focus on hash address $\queryhash$, the address with $t$ (all) $0$'s. The hash addresses that are hamming distance $r$ from $\queryhash$ contain exactly $r$ $1$'s.

We first take $\sum_{j \in \neighbors} \countsvectornotablej[r-1]$, the sum of the elements that are hamming distance $r-1$ from $\queryhash$'s immediate neighbors. However, this sum also includes elements that are $r-1$ distance from $\queryhash$'s neighbors via $\queryhash$ as an intermediate hop, which consequently are not distance $r$ from $\queryhash$, but rather $r-2$ distance from $\queryhash$, since each neighbor is distance 1 from $\queryhash$. We claim that there are exactly $(t-r+2)C_{\queryhash}[r-2]$ of these elements that were included in the sum. Fix any hash address $h$ that is distance $r-2$ from $\queryhash$. Clearly $h$ has exactly $(r-2)$ 1's in its address. There are $(t-r+2)$ 0's in $h$'s address. Any neighbor $j$ of $\queryhash$ that has a 1 in any of those $t-r+2$ slots will report $h$ as part of its count $\countsvectornotablej[r-1]$. There are $t-r+2$ such neighbors, therefore, we must adjust $\sum_{j \in \neighbors} \countsvectornotablej[r-1]$ by subtracting $(t-r+2)\countsvectornotable[r-2]$. 

There is one other source of double-counting, which is that many neighbors $j$ of $\queryhash$ will include the same bucket of hamming distance $r$ away from $\queryhash$ as part of their $\countsvectornotablej[r-1]$ count. This over-counting can be quantified in the following way. Fix any hash address $d$ that is distance $r$ from $\queryhash$. Clearly the sequence $d$ contains exactly $(r)$ $1$'s. So any neighbor $j$ of $\queryhash$ that contains a $1$ in any one of those $r$ slots will include $d$ as part of its $\countsvectornotablej[r-1]$  count, and there are $r$ such neighbors. So our final expression for $C_{\queryhash}(r)$ is 
\[
C_{\queryhash}[r] = \frac{\sum_{j \in \neighbors} \countsvectornotablej[r-1] - (t-r+2)C_{\queryhash}[r-2]}{r}
\]

This same argument generalizes to any bucket $i$, so we conclude that our general update rule $\countsvectornotable[r] = \frac{\sum_{j \in \neighbors} \countsvectornotablej[r-1] - (t-r+2)\countsvectornotable[r-2]}{r}$ is correct. 

$\aggregatecounts$ proceeds in $t+1$ rounds in the outer loop. Each round iterates over $2^t$ buckets in the inner loop. An update for each bucket $i$ looks at the neighbors $\neighbors$ of $i$, and each hash address $i$ has exactly $t$ neighbors (each corresponding to one bit flip in the length $t$ hash address). So $\aggregatecounts$ terminates in time $O(t^2 2^t)$. 

When we want to compute the counts vectors for many tables, we invoke $\aggregatecounts$ for each of $K$ tables so the overall runtime is $O(Kt^2 2^t)$.

\section{Constructing a Uniform Sampler for Fixed Hamming Distances across Multiple Tables} \label{appendix_B}
We construct a sampler that, given a different hash address $i^k$ for each table $k \in [K]$, and a certain set of hamming distances $\mathcal{I}$, returns a data point uniformly at random from the union of the set of elements in each table that are contained in buckets of hamming distance $\mathcal{I}$ from hash address $i^k$. 

The high level implementation of our sampler works as follows. Given a set of hamming distances $\mathcal{I}$ that we are interested in and hash address $i^k$ for each of our $K$ tables, we know from Appendix A that we can construct $\countsvector$ that gives the number of elements in buckets that are hamming distance $0, 1, \ldots t$ away from address $i^k$ for each table $k$. Then for each table, we can add the relevant indices to obtain the total count for elements at hamming distances $\mathcal{I}$, that is, we can compute $\sum_{d \in \mathcal{I}} \countsvector[d]$ for each $k$. We choose to take a sample from table $k^*$ with probability $\frac{\sum_{d \in \mathcal{I}} C^{k^*}_{i}[d]}{\sum_k \sum_{d \in \mathcal{I}} C^{k}_{i}[d]}$.

Now that we have fixed our choice of table $k^*$, we want to pick a particular hamming distance within $\mathcal{I}$ to sample from. This can be done using the counts vector for that table, in particular we choose the hamming distance $d^* \in \mathcal{I}$ with probability $\frac{C^{k^*}_i[d^*]}{\sum_{d \in \mathcal{I}} C^{k^*}_i[d]}$. 

Having now fixed a table $k^*$ and a particular hamming distance $d^*$, we introduce an algorithm $\hammingdistancesampler$ that generates a sample uniformly from the set of elements hashed to buckets at hamming distance $d^*$ from address $\queryhash$ (without loss of generality) in table $k^*$. Our algorithm uses the counts matrix $\countsmatrix$ as the underlying data structure. 
Our main results says: 
\sampler*

We first describe the implementation of the sampler in Section $\ref{ssec:uniform_sampler}$ and then later describe the implementation for constructing the Counts Matrix $\countsmatrix$ in Section $\ref{ssec:counts_matrix}$. Our main result follows from Lemma $\ref{lem:sampler}$ and Lemma $\ref{lem:counts_matrix}$

\subsection{Uniform Sampler for One Table}
\label{ssec:uniform_sampler}
In this section, we describe our algorithm, $\hammingdistancesampler$ (Algorithm \ref{alg:hamming_distance_sampler}), which helps to generate a sample uniformly at random, from one hash table, an element from hamming distance $d$ to hash address $i$. Suppose for each hash address $i$ in the table we are given its counts matrix $\countsmatrixnotable \in \Z^{(t+1) \times (t+1)}$ such that $\countsmatrixnotable[s,a]$ gives a count of the number of elements hashed to buckets whose addresses are at hamming distance $s$ away from $i$ and share the same first $a$ bits as address $i$. $\hammingdistancesampler$ uses this counts matrix to help decide which hash bucket to sample from. 

$\hammingdistancesampler$ (Algorithm \ref{alg:hamming_distance_sampler}) chooses a target hash address that is hamming distance $d$ from $\queryhash$ by iteratively generating a bit pattern to XOR with the query hash address. Without loss of generality, suppose the query hash address is $\queryhash$. We start from left to right. We set the first bit of the XOR mask to 1 with probability proportional to the number of elements at hamming distance $d$ to $\queryhash$ that have 1 as their first bit. 

Now that we have decided on the first bit of the XOR mask, we move on to the second bit. Conditioned on our choice for the first bit, we make our second choice. If we had chosen 1 for the first bit of the mask, now we choose to set the second bit of the XOR mask with probability proportional to the number of elements at hamming distance $d$ to $\queryhash$ that have 11 as their first two bits, and we choose to set the second bit to 0 with probability proportional to the number of elements at hamming distance $d$ to $\queryhash$ that have 10 as their first two bits, and so on and so forth. 

We continue until we arrive at a target hash address that is exactly hamming distance $d$ from $i$, which is the output of $\hammingdistancesampler$. After we choose our target sampling hash address, we sample uniformly at random the elements within that hash bucket. $\hammingdistancesampler$ is formally written as Algorithm \ref{alg:hamming_distance_sampler}.
 
\begin{algorithm}[ht]
\caption{Hamming Distance Sampler}
\label{alg:hamming_distance_sampler}
\begin{algorithmic}[1]
\REQUIRE{Hash table with buckets $B$, hash address $i$, hamming distance $d$, counts matrix $M_b$ for all buckets $b \in B$}
\STATE $\text{MASK} = \queryhash$
\STATE $g = 0$
\FOR{round $r=0, r\text{++}, r < t$}
\IF{$g < d$}
\STATE $i' = i \oplus \text{MASK}$
\STATE $i'' = i \oplus \text{MASK} \oplus {\bf{0}}_r{\bf{1}}{\bf{0}}_{t-r-1}$\vspace*{0.1cm}
\STATE $p_{r} = \frac{M_{i''}[d-g-1, r+1]}{M_{i''}[d-g-1, r+1] + M_{i'}[d-g, r+1]}$
\STATE Flip a biased coin with probability $p_r$ of coming up heads. Let $f = 1$ if heads, $f = 0$ else. 
\IF{$f = 1$}
\STATE $g \leftarrow g + 1$ \hspace*{1cm} //Update the count of 1's already chosen
\STATE $\text{MASK} \leftarrow \text{MASK} \oplus {\bf{0}}_r{\bf{1}}{\bf{0}}_{t-r-1}$ \hspace*{1cm} //Update the $\text{MASK}$ to make the $(r+1)$-th bit 1
\ENDIF
\ENDIF
\ENDFOR
\STATE \textbf{Return} $i \oplus \text{MASK}$
\end{algorithmic}
\end{algorithm}

\subsubsection{Analysis of Algorithm}

We prove the following guarantee for $\hammingdistancesampler$ (Algorithm \ref{alg:hamming_distance_sampler}): 

\begin{lemma}
\label{lem:sampler}
Suppose there are a total of D elements that are contained in buckets of hamming distance $d$ from hash address $i \in \{0,1\}^t$, and bucket $b$ which is hamming distance $d$ from address $i$ contains $m$ elements. Then $\hammingdistancesampler$ (Algorithm \ref{alg:hamming_distance_sampler}) returns address $b$ with probability $\frac{m}{D}$ in time $O(t)$. 
\end{lemma}

Once we have the output of $\hammingdistancesampler$, which is a hash address $b$ that was generated with probability $\frac{m}{D}$, then we can pick an element uniformly at random from within bucket $b$ to generate a sample with uniform probability $\frac{1}{D}$. 

\begin{proof}

Without loss of generality, we assume that the hash address of interest $i = \queryhash$. 

We first notice that the set of all possible realizations of this algorithm can be represented as a binary tree with depth at most $t$, and each round can be viewed as traversing the binary tree. We first describe this tree. The root of the tree has label $\queryhash$, and its value is the total number of elements in buckets at hamming distance $d$ from $\queryhash$. Its left child node has label $\queryhash$ and its value is the total number of elements across all buckets at hamming distance $d$ away from $\queryhash$ that share the first bit (0). The root's right child node has label ${\bf{1}}{\bf{0}}_{t-1}$ and its value is the total number of elements across all buckets that are hamming distance $d-1$ away from ${\bf{1}}{\bf{0}}_{t-1}$. 

In general, each node $V$ at depth $r$ can be expressed as a label, value pair  $(l, v)$, where the label $l$ is a hash address, and if we let $g = d(i, l)$, and the value $v$ is a count of elements at hamming distance $d-g$ away from the label hash address $l$. Its left child is labeled by $l$ and its value is the total number of elements across all buckets at hamming distance $d-g$ away from the label of the node (l) that match the first $r+1$ bits as its label. The parent's right child is labeled by $l \oplus {\bf{0}}_r{\bf{1}}{\bf{0}}_{t-r-1}$. and its value is the total number of elements across all buckets that are hamming distance $d-g-1$ away from its label and that share the first $r+1$ bits as its label.

Clearly the leaves of the tree are the set of labels (bucket addresses) that are of hamming distance $d$ away from $i$, and have values that correspond to the number of elements in each bucket. We also note that the label of each node corresponds to the XOR mask in our algorithm, and the value corresponds to an entry in the $\countsmatrix$.

We now analyze the correctness of $\hammingdistancesampler$ using proof by induction. We start at the root of the tree, with label ${\bf{0}}_t$. We want to show that the probabilities that the XOR mask takes on a specific value at the end of round $r$ for $r = 0, \ldots t$ lead to uniform probabilities of choosing an element hashed to a bucket at hamming distance $d$ from ${\bf{0}}_t$.

Suppose further there are a total of D elements that are contained in buckets of hamming distance $d$ from hash address ${\bf{0}}_t$. \\

\textbf{Base Case:} After round $r=0$, the probability that the XOR mask becomes ${\bf{1}}{\bf{0}}_{t-1}$ is $\frac{M_{{\bf{1}}{\bf{0}}_{t-1}}[d-1, 1]}{M_{{\bf{0}}_{t}}[d, 1] + M_{{\bf{1}}{\bf{0}}_{t-1}}[d-1, 1]} = \frac{M_{{\bf{1}}{\bf{0}}_{t-1}}[d-1, 1]}{D}$. \\

\textbf{Inductive Hypothesis:} Assume that at end of round $r-1$, the probability that we reach a certain node at level $r$ with label (XOR mask) $l$, and with $g_l$ 1's in the mask, is $\frac{M_{l}[d-{g_l}, r]}{D}$. \\

\textbf{Inductive Step:} We want to show that at the end of round $r$, the probability that we reach a certain node at level $r+1$ with label $c$ (for child), and with $g_c$ 1's in the mask, is $\frac{M_{c}[d-{g_c}, r+1]}{D}$. 

This follows directly from the inductive hypothesis. Note that at the end of round $r-1$, we have reached a certain node at level $r$ with label (XOR mask) $l$, and with $g_l$ 1's in the mask, is $\frac{M_{l}[d-{g_l}, r]}{D}$. From this node, the probability of reaching the left child $lc$ is  $\frac{M_{lc}[d-{g_{lc}}, r+1]}{M_{l}[d-{g_l}, r]}$ and the probability of reaching the right child is $\frac{M_{rc}[d-{g_{rc}}, r+1]}{M_{l}[d-{g_l}, r]}$. Multiplying this by the probability that we reach the parent from the inductive hypothesis gives the proof of the inductive step.

Since there are a total of D elements that are contained in buckets of hamming distance $d$ from hash address $\queryhash$, and bucket $b$, which is hamming distance $d$ from address $i$, contains $m$ elements. The probability of reaching $b$ by traversing down this tree is $\frac{m}{D}$. 

Additionally, this algorithm takes $O(t)$ time to produce the hash address $b$. This is clear since the tree has depth at most $t$, and each step in the traversal is a constant time operation.
\end{proof}

\subsection{Computing the Counts Matrix}
\label{ssec:counts_matrix}
Since $\hammingdistancesampler$ (Algorithm \ref{alg:hamming_distance_sampler}) requires a matrix $\countsmatrixnotable \in \Z^{(t+1) \times (t+1)}$ such that $\countsmatrixnotable[s,a]$ gives a count of the number of elements that are hamming distance $s$ away from $i$ and share the same first $a$ bits as address $i$, we show how to precompute such a counts matrix in time $O(K t^3 2^t)$ for each $i \in \{0,1\}^t$ over the $K$ hash tables. 

\begin{lemma}
\label{lem:counts_matrix}
The counts matrix $\countsmatrix \in \Z^{(t+1) \times (t+1)}$ can be computed in time $O(K t^3 2^t)$ for each $i \in \{0,1\}^t$ and $k \in \{1, \ldots K\}$. 
\end{lemma}

\begin{proof}
We use the algorithm that we develop in Appendix A, $\aggregatecounts$ (Algorithm ~\ref{alg:aggregate_counts}),  to compute matrix $\countsmatrixnotable$. 

Now, to use $\aggregatecounts$, we observe that $\countsmatrixnotable[s, a]$ for all $s \in \{0, 1, \ldots t\}$ is just another instance of the $\aggregatecounts$ problem, restricted to the case where we only consider buckets that match $i$ on the first $a$ bits. Clearly, by the Appendix A Main Theorem we can compute $\countsmatrixnotable[s][0]$ for all $s \in \{0, \ldots t\}$ (the entire column) simultaneously for all $i$ in time $O(t^2 2^t)$. 

In fact, fixing each $a \in \{0, 1, \ldots t\}$, it is possible to compute $\countsmatrixnotable[s][a]$ for all $s \in \{0, \ldots t\}$ (the entire column) simultaneously for all $i$ in time $O(t^2 2^t)$. This is because in our updates for each $i$, we can just consider the buckets that match $i$ on the first $a$ bits. We can invoke $\aggregatecounts$ using neighbor buckets $\neighbors(a)$, and the number of rounds would be $t - a$. The runtime to update each column of $\countsmatrixnotable$ (fixing $a$ and over all $i$) is $O(t^2 2^t)$ (one invocation of $\aggregatecounts$), so the total runtime to compute $\countsmatrix$ over all $i$ is $O(t^3 2^t)$. Repeating for each of $K$ tables yields the final runtime of $O(K t^3 2^t)$.
\end{proof}
\section{Multi-Probe for Random Hyperplane LSH} \label{appendix_C}

We specialize the multi-probe technique introduced in \cite{lv2007multi} for random hyperplane LSH. First, we briefly overview the details of random hyperplane LSH \cite{Charikar:2002:SET:509907.509965}. In this scheme, each hash vector is chosen from the $d$-dimensional Gaussian distribution (each coordinate is drawn from the 1-dimensional Gaussian distribution). Each hash vector $r$ contributes one bit to the hash sequence of a data point $v$, based on the rule: 
\begin{equation}\label{eq:simhash}
h_r(v) = \begin{cases} 
      0 & \text{ if } r \cdot v \leq 0 \\
     1 & \text{ otherwise. }
   \end{cases}
\end{equation}
The following 2 facts are well-known: 
\begin{itemize}
\item For any query vector $q$ and data vector $x$, 
\[
\P\left(h_r(q) \neq h_r(x)\right) =  \frac{\theta_{qx}}{\pi} ~.
\]
\item Moreover, for any hamming distance $i$,
\[
\P(d_{qx} = i | \theta_{qx}) = {t \choose i} \left(1 - \frac{\theta_{qx}}{\pi}\right)^{t-i} \left(\frac{\theta_{qx}}{\pi}\right)^i ~,
\]

where $d_{qx}$ is the hamming distance between the hash for query $q$ and the hash for data vector $x$, $\theta_{qx}$ denotes the angle between the 2 vectors, and $t$ is the total number of bits in the hash sequence. 
\end{itemize}

Multi-probe is a strategy for probing multiple buckets in a hash table to leverage the fact that while the bucket that the query hashes to has the highest probability of containing interesting data points, nearby buckets can also be very useful.\cite{lv2007multi} introduces two multi-probe methods, step-wise (probe buckets at $0, 1, 2, \ldots$ hamming distance away), and query-directed (calculate collision probabilities for specific hash addresses based on the query's exact projection onto the hash functions $r$, and then probe the most promising buckets).

We focus on query-directed multi-probe, as  \cite{lv2007multi}  shows is more effective than step-wise multi-probe. We provide the first analysis for the exact collision probabilities and other necessary implementation details for query-directed multi-probe using random hyperplane LSH. Our main contribution is the following theorem. 
\begin{theorem}[Multi-probe Collision Probability]
Let $r$ be a random hyperplane drawn from the $d$-dimensional Gaussian distribution, and for an arbitrary vector $v \in \R^d$, let $h_r(v)$ be defined as in Equation \ref{eq:simhash}. Moreover, for a query vector $q \in \R^d$, let $r_q$ be the value of the inner product $r \cdot q$. Then $\P\left(h_r(x) \neq h_r(q) | r_q \right) = \frac{1}{2} - \frac{1}{2} \erf{\left(\frac{|r_q|}{\sqrt{2} \tan(\theta_{qx})}\right)}$, where $\erf$ denotes the error function.
\end{theorem}

\subsection{Preliminaries}
Let $q$ be a query vector and $x$ be a data vector in $\R^d$, normalized to unit length. Suppose that $r$ is a random hyperplane hash vector drawn from the $d$-dimensional Gaussian distribution. In our problem setting, $x$ has already been assigned to a bucket $b$ in a hash table as part of a preprocessing step, so we know $h_r(x)$ as a binary 0-1 value. However, since we compute the projection of $q$ onto $r$ in an on-line fashion, we have access to $q \cdot r$ (not just the binary $h_r(q)$ value). 

Let $q \cdot r$ be denoted $r_q$. Then our goal is to compute $\P\left(h_r(x) \neq h_r(q) | r_q \right)$. To compute this probability, a few observations will be useful. First, it is helpful to decompose $x$ and $r$ in terms of $q$. Let $q^{\perp}$ denote an arbitrary vector in $\R^d$ satisfying $q \cdot q^{\perp} = 0$. Then we can rewrite $x$ and $r$ as: 
$$ x = \cos(\theta_{qx}) \cdot q + \sin(\theta_{qx})\cdot q^{\perp}$$
$$ r = r_q \cdot q + r_{q^{\perp}}\cdot q^{\perp}$$
Moreover, we know $r_q$ and $r_{q^{\perp}}$. 

Another helpful fact is that by the 2-stability of the Gaussian distribution, $r_q$ and $r_{q^{\perp}}$ are both standard normal random variables. 
\subsection{Collision Probabilities}
Now we focus our attention on $\P\left(h_r(x) \neq h_r(q) | r_q \right)$. First, we remark that if we know the value of $r_q$, then we also know $h_r(q)$. Now to analyze $h_r(x)$, we need to analyze $x \cdot r$. We can write:
\begin{align*}
x \cdot r &= \langle \cos(\theta_{qx}) \cdot q + \sin(\theta_{qx})\cdot q^{\perp}, r_q \cdot q + r_{q^{\perp}}\cdot q^{\perp} \rangle \\
& =  r_q \cos(\theta_{qx}) + r_{q^{\perp}}\sin(\theta_{qx})
\end{align*}
So we consider the two cases where $h_r(q) = 0$ and $h_r(q) = 1$. Note that these cases are equivalent to $r_q \leq 0$ and $r_q > 0$, respectively. \\
Under $h_r(q) = 0$, we have: 
\begin{align*}
\P\left(h_r(x) \neq h_r(q) | r_q \right) =& \P \left(r_{q} \cos(\theta_{qx}) + r_{q^{\perp}}\sin(\theta_{qx}) > 0 | r_q\right)\\
= & \P \left(r_{q^{\perp}}\sin(\theta_{qx}) > |r_{q}| \cos(\theta_{qx}) | r_q\right)\\
= & \P \left(r_{q^{\perp}} > |r_{q}| \frac{\cos(\theta_{qx})}{\sin(\theta_{qx})} | r_q\right) \\
= & \P \left(r_{q^{\perp}} > |r_{q}| \frac{1}{\tan(\theta_{qx})} | r_q\right) \\
= & 1- \P\left(r_{q^{\perp}} < |r_{q}| \frac{1}{\tan(\theta_{qx})} | r_q\right) \\
= & \frac{1}{2} - \frac{1}{2} \erf{\left(\frac{|r_q|}{\sqrt{2} \tan(\theta_{qx})}\right)}~.
\end{align*}
Under $h_r(q) = 1$, we have: 
\begin{align*}
\P\left(h_r(x) \neq h_r(q) | r_q\right) =& \P \left(r_{q} \cos(\theta_{qx}) + r_{q^{\perp}}\sin(\theta_{qx}) < 0 | r_q\right)\\
= & \P \left(r_{q^{\perp}}\sin(\theta_{qx}) < -|r_{q}| \cos(\theta_{qx}) | r_q \right)\\
= & \P \left(r_{q^{\perp}} < -|r_{q}| \frac{\cos(\theta_{qx})}{\sin(\theta_{qx})} | r_q\right)\\
= & \P \left(r_{q^{\perp}} < -|r_{q}| \frac{1}{\tan(\theta_{qx})} | r_q \right) \\
= & \frac{1}{2} - \frac{1}{2} \erf{\left(\frac{|r_q|}{\sqrt{2} \tan(\theta_{qx})}\right)} ~.
\end{align*}
Therefore, we conclude that 
$$\P\left(h_r(x) \neq h_r(q) | r_q \right) = \frac{1}{2} - \frac{1}{2} \erf{\left(\frac{|r_q|}{\sqrt{2} \tan(\theta_{qx})}\right)}~.$$
\subsection{Probing Sequence}
Clearly, for a hash sequence that appends the hash values from multiple random hyperplanes, one can use the above formula to compute the exact probabilities of $x$ landing in a particular bucket address relative to the query $q$, given $\theta_{qx}$ and $r_q$ for each $r$. Suppose $p_i = \P\left(h_{r^i}(x) \neq h_{r^i}(q) | r^i_q \right)$ denote the probability that the $i$-th bit corresponding to the $i$-th random hyperplane projection is flipped. 

Let $g$ denote the hash function for the table, which is a composite function over a set of random hyperplanes. Then the probability that $x$ hashes to bucket $b$ (the event that $g(x) = b$) given the query bucket, $g(q)$, and the query's projections onto the hyperplanes, denoted $r_q$, is: 
$$\P\left(g(x) = b | r_q, g(q)\right) = \underset{i \in \text{bits flipped}} \prod (p_i) \underset{j \in \text{bits unflipped}} \prod (1-p_j)~,$$
where the sets of bits flipped and bits unflipped are with respect to the hash sequences $g(x) = b$ and $g(q)$. 
\section{The Joint Probability of a Random Hyperplane Cut: An Extension of Goemans-Williamson for 3 vectors}\label{appendix_D}

We extend the Goemans-Williamson random hyperplane cut analysis\cite{goemans1995improved} for a system of 3 vectors, query vector $q$, and data vectors $x$ and $y$ in $\R^d$, normalized to unit length. In particular, we are interested in studying $$p(x,y) = \P(d_{qx} \in \mathcal{I} \cap d_{qy} \in \mathcal{I}| \theta_{qx}, \theta_{qy}, \theta_{xy})~,$$ the collision probability that $x$ and $y$ both land in buckets that are hamming distance $\mathcal{I}$ from $q$ over the random choice of hash functions. Let $\theta_{qx}, \theta_{qy}, \theta_{xy}$ denote the angles between $q$ and $x$, $q$ and $y$, and $x$ and $y$, respectively. 

In general, $p(x,y) \leq \min\{p(x), p(y)\}$. The worst case 3-vector configuration achieves $p(x,y) =  p(x)$. However, for many configurations a finer analysis yields a more exact probability. In this appendix, we prove the following main theorem:
\begin{theorem}
\label{thm:gw_extension}
Let $t$ be the total number of hash bits in the hash sequence. Let $\mathcal{I}$ be a set of hamming distances, and define $p(x,y) = \P(d_{qx} \in \mathcal{I} \cap d_{qy} \in \mathcal{I}| \theta_{qx}, \theta_{qy}, \theta_{xy})$. Then 
$$p(x,y) = \sum\limits_{a,b \in \mathcal{I}} \P(d_{qx} = a \cap d_{qy} = b | \theta_{qx}, \theta_{qy}, \theta_{xy})$$
where 
\begin{align*}
\P(d_{qx} = a \cap d_{qy} = b | \theta_{qx}, \theta_{qy}, \theta_{xy}) = \sum\limits_{i=0}^a &{t \choose i} \left[\frac{\theta_{qx} + \theta_{qy} - \theta_{xy}}{2\pi}\right]^i \\
& \cdot {t-i \choose b-i} \left[ \frac{-\theta_{qx} + \theta_{qy} + \theta_{xy}}{2\pi} \right]^{b-i} \\
& \cdot {t-b \choose a-i} \left[\frac{\theta_{qx} - \theta_{qy} + \theta_{xy}}{2\pi} \right]^{a-i} \\
& \cdot \left[1 - \frac{\theta_{qx} + \theta_{qy} + \theta_{xy}}{2\pi}\right]^{t-a-b+i}~.
\end{align*}
\end{theorem} 
To show Theorem \ref{thm:gw_extension}, we first analyze $h_r(\cdot)$ for a 3 vector configuration, $x, y,$ and $q$. The building blocks of the analysis are the following quantities, for which we will individually derive expressions for: 

$$\P\left(h_r(q) = h_r(x) \cap h_r(q) = h_r(y)\right)$$
$$\P\left(h_r(q) = h_r(x) \cap h_r(q) \neq h_r(y)\right)$$
$$\P\left(h_r(q) \neq h_r(x) \cap h_r(q) = h_r(y)\right)$$
$$\P\left(h_r(q) \neq h_r(x) \cap h_r(q) \neq h_r(y)\right)$$

\subsection{Analysis}
We start by analyzing the probability that $r$ cuts between $q$ and $x$, as well as $q$ and $y$. 
\begin{lemma}
$$\P\left(h_r(q) \neq h_r(x) \cap h_r(q) \neq h_r(y)\right) = \frac{\theta_{qx} + \theta_{qy} - \theta_{xy}}{2\pi}~.$$
$$\P\left(h_r(q) = h_r(x) \cap h_r(q) \neq h_r(y)\right) = \frac{-\theta_{qx} + \theta_{qy} + \theta_{xy}}{2\pi}~.$$
$$\P\left(h_r(q) \neq h_r(x) \cap h_r(q) = h_r(y)\right) = \frac{\theta_{qx} - \theta_{qy} + \theta_{xy}}{2\pi}~.$$
$$\P\left(h_r(q) = h_r(x) \cap h_r(q) = h_r(y)\right) = 1 - \frac{\theta_{qx} + \theta_{qy} + \theta_{xy}}{2\pi}~.$$
\end{lemma}
\begin{proof}
We focus first on $\P\left(h_r(q) \neq h_r(x) \cap h_r(q) \neq h_r(y)\right).$
The following three expressions follow immediately from the law of total probability: 
\begin{align}
\P\left(h_r(q) \neq h_r(x) \cap h_r(q) \neq h_r(y)\right) + \P\left(h_r(q) \neq h_r(x) \cap h_r(q) = h_r(y)\right) &= \P(h_r(q) \neq h_r(x))\notag\\
&= \frac{\theta_{qx}}{\pi}\label{qx} \\\notag\\
\P\left(h_r(q) \neq h_r(x) \cap h_r(q) \neq h_r(y)\right) + \P\left(h_r(q) = h_r(x) \cap h_r(q) \neq h_r(y)\right) &= \P(h_r(q) \neq h_r(y))\notag\\
&= \frac{\theta_{qy}}{\pi}\label{qy} \\\notag\\
\P\left(h_r(q) \neq h_r(x) \cap h_r(q) = h_r(y)\right) +\P\left(h_r(q) = h_r(x) \cap h_r(q) \neq h_r(y)\right) &= \P(h_r(x) \neq h_r(y))\notag\\
&= \frac{\theta_{xy}}{\pi}\label{xy} \\\notag
\end{align}
Adding \ref{qx} and \ref{qy} and subtracting \ref{xy} yields the result. A similar analysis can be used to derive the other expressions. 
\end{proof}
Therefore, for $a, b \in \mathcal{I}$, where without loss of generality, $a \leq b$, we can say that: 
\begin{align*}
\P(d_{qx} = a \cap d_{qy} = b | \theta_{qx}, \theta_{qy}, \theta_{xy}) = \sum\limits_{i=0}^a &{t \choose i} \left[\P\left(h_r(q) \neq h_r(x) \cap h_r(q) \neq h_r(y)\right)\right]^i \\
& \cdot {t-i \choose b-i} \left[ \P\left(h_r(q) = h_r(x) \cap h_r(q) \neq h_r(y)\right) \right]^{b-i} \\
& \cdot {t-b \choose a-i} \left[\P\left(h_r(q) \neq h_r(x) \cap h_r(q) = h_r(y)\right) \right]^{a-i} \\
& \cdot \left[\P\left(h_r(q) = h_r(x) \cap h_r(q) = h_r(y)\right)\right]^{t-a-b+i} ~.
\end{align*}
One can then substitute the appropriate quantities from above. 
Finally, we can evaluate $p(x,y)$ as:
$$p(x,y) = \P(d_{qx} \in \mathcal{I} \cap d_{qy} \in \mathcal{I}| \theta_{qx}, \theta_{qy}, \theta_{xy}) = \sum\limits_{a,b \in \mathcal{I}} \P(d_{qx} = a \cap d_{qy} = b | \theta_{qx}, \theta_{qy}, \theta_{xy})~.$$
This concludes the proof of Theorem \ref{thm:gw_extension}.

\end{document}